\documentstyle[emulateapj,apjfonts,epsfig]{article}

\renewcommand{\vec}[1]{\mbox{\boldmath $\displaystyle #1$}} 
\newcommand{\grad}{\vec{\nabla}}   

\newcommand{\edd}{_{\rm Edd}}
\newcommand{\zcno}{Z_{\rm CNO}}
\newcommand{\BV}{Brunt-V\"ais\"al\"a }

\begin{document}

\submitted{To appear in The Astrophysical Journal}

\title{Rotational Evolution during Type I X-Ray Bursts}

\author{Andrew Cumming\altaffilmark{1}}
\affil{Department of Physics, University of California, Berkeley, CA 94720 \\ 
cumming@itp.ucsb.edu\\}
\vspace{0.2cm}
\author{Lars Bildsten} 
\affil{Institute for Theoretical Physics and Department of Physics\\
Kohn Hall, University of California, Santa Barbara, CA 93106 \\ bildsten@itp.ucsb.edu\\}

\altaffiltext{1}{Present address: Institute for Theoretical Physics,
Kohn Hall, University of California, Santa Barbara, CA 93106.}

\begin{abstract}

The rotation rates of six weakly-magnetic neutron stars accreting in
low-mass X-ray binaries have most likely been measured by Type I X-ray
burst observations with the {\it Rossi X-Ray Timing Explorer}
Proportional Counter Array.  The phenomenology of the nearly coherent
oscillations detected during the few seconds of thermonuclear burning is
most simply understood as rotational modulation of brightness
asymmetries on the neutron star surface. We show that, as suggested by
Strohmayer and colleagues, the frequency changes of 1--2 Hz observed
during bursts are consistent with angular momentum conservation as the
burning shell hydrostatically expands and contracts during the burst. We
calculate how vertical heat propagation through the radiative outer
layers of the atmosphere and convection affect the coherence of the
oscillation.  We show that the evolution and coherence of the rotational
profile depends strongly on whether the burning layers are composed of
pure helium or mixed hydrogen/helium.  Our results help explain the
absence (presence) of oscillations from hydrogen-burning (helium-rich)
bursts that was found by Muno and collaborators.

We also investigate angular momentum transport within the burning layers
and address the recoupling of the burning layers with the star. We show
that the Kelvin-Helmholtz instability is quenched by the strong
stratification, and that mixing between the burning fuel and underlying
ashes by the baroclinic instability does not occur. However, the
baroclinic instability may have time to operate within the
differentially rotating burning layer, potentially bringing it into
rigid rotation.
\end{abstract}



\section{Introduction}\label{sec:Intro}

Type I X-ray bursts have been long understood as thermonuclear flashes
on the surfaces of neutron stars accreting at rates  of $10^{-11}
M_\odot \ {\rm yr^{-1}} < \dot M < 10^{-8} M_\odot \ {\rm yr^{-1}}$ 
in low mass X-ray binaries (LMXBs). The accreted hydrogen and helium
accumulates on the surface of the neutron star and periodically
ignites and burns. Thermonuclear flash models successfully explain the
burst recurrence times (hours to days), energetics ($\sim
10^{39}\,{\rm ergs}$), and durations ($\sim 10$--$100\,{\rm s}$)
(Lewin, van Paradijs, \& Taam 1993; Bildsten 1998), though many
quantitative comparisons to observations are less successful (for
example, see Fujimoto et al. 1987; Bildsten 2000).

\begin{deluxetable}{llccccccl}
\tiny
\footnotesize
\tablecaption{Properties of Observed Type I Burst Oscillations\label{tab:obs}}
\tablewidth{0pt}
\tablehead{
\colhead{Object} & \colhead{Time} & \colhead{Radius} &
\colhead{$\nu_0$} & \colhead{$\Delta\nu$} &
\colhead{$\tau$} & \colhead{$\Delta\nu/\nu_0$} & 
\colhead{Oscillations} & \colhead{References} \\
\colhead{} & \colhead{(UT)} & \colhead{Expansion?} &
\colhead{(Hz)} & \colhead{(Hz)} &
\colhead{(s)} & \colhead{($10^{-3}$)} & \colhead{During Rise?} & \colhead{}
}
\startdata
4U 1636-54  & 1996 Dec 28 (22:39:22) & Y & 580.5 & $\approx 1$--$2.5$
& \nodata & $\approx 2$--$4$ &Y &1,2\tablenotemark{b}\\
            & 1996 Dec 29 (23:26:46) & Y& 581.5 & $\approx 2$
& \nodata & $\approx 3$ &Y &2,3\tablenotemark{b}\\
            & 1996 Dec 31 (17:36:52)\tablenotemark{c} & Y& 581 & $\approx 2$
& \nodata & $\approx 3$ &Y &2,3\tablenotemark{b}\\
4U 1702-43  & 1997 Jul 26 (14:04:19) & ? & $329.85\pm 0.1$ & 2.5 &
$1.88\pm 0.25$ & $7.7\pm 0.3$ & Y & 4\tablenotemark{a}\\
            & 1997 Jul 30 (12:11:58) & ? & $330.55\pm 0.02$ & 1.6 &
$4.02\pm 0.07$ & $4.8\pm 0.3$ & N & 4\tablenotemark{a}\\
4U 1728-34  & 1996 Feb 16 (10:00:45) & N & $364.23\pm 0.05$ & 2.4 &
$3.52\pm 0.28$ & $6.6\pm 0.1$ & Y &4,5\tablenotemark{a}\\
            & 1997 Sep 9 (06:42:56) & ? & $364.10\pm 0.05$ & 2.1 &
$1.84\pm 0.15$ & $5.9\pm 0.2$ & ? &4\tablenotemark{a}\\
KS 1731-26 & 1996 Jul 14 (04:23:42) & Y & $524.61^{+0.13}_{-0.07}$ &
$0.96\pm 0.04$ & $4.1^{+1.3}_{-0.9}$ & 1.8 & N &6,7\tablenotemark{a}\\
& 1999 Feb 27 (17:23:01) & Y & $524.48^{+0.05}_{-0.03}$ &
$1.06\pm 0.05$ & $2.6^{+0.4}_{-0.3}$ & 2.0 &N & 7\tablenotemark{a}\\
Galactic Center & 1996 May 15 (19:32:10) &Y & 589 & $\approx 1$ & \nodata &
$\approx 2$&
N &8\tablenotemark{b} \\
(MXB 1743-29?) &&&&&&&&\\
Aql X-1 & 1997 Mar 1 (23:27:39) & N & $549.76\pm 0.04$ &
$2.38^{+0.03}_{-0.07}$ & $2.69\pm 0.15$ & 4.3 & N & 9,10\tablenotemark{a}\\
\enddata
\tablenotetext{a}{In these cases, the frequency evolution was fitted
by an exponential model $\nu(t)=\nu_0-\Delta\nu e^{-t/\tau}$.}
\tablenotetext{b}{For these bursts, $\nu_0$ is the frequency seen
in the tail, and $\Delta\nu$ how much the frequency changes during the
burst.}
\tablenotetext{c}{This burst observed an episode of spin down in the
tail (Strohmayer 1999b).}
\tablenotetext{}{References.---(1)
Strohmayer et al. 1998c; (2) Miller 2000; (3) Strohmayer 1999a; (4)
Strohmayer \& Markwardt 1999; (5) Strohmayer et al. 1996; (6) Smith,
Morgan \& Bradt 1997; (7) Muno et al.  2000; (8) Strohmayer et
al. 1997a; (9) Zhang et al. 1998; (10) Fox et al. 2000.}
\end{deluxetable}

Evolutionary scenarios connecting the neutron stars in LMXBs to the
millisecond radio pulsars (see Bhattacharya 1995 for a review) predict
that neutron stars in LMXBs should be spinning rapidly. This has been
confirmed for one system, the $\nu_s=401$ Hz accreting pulsar SAX
J1808.4-3658 (Wijnands \& van der Klis 1998; Chakrabarty \& Morgan
1998), in which the neutron star magnetic field ($B\sim 10^8$--$10^9\
{\rm G}$; Psaltis \& Chakrabarty 1999) channels the accretion flow onto
the magnetic polar caps, creating an asymmetry which is modulated by
rotation. However, most neutron stars in LMXB's show no evidence for
coherent periodicity in the persistent emission, implying that the
neutron stars do not possess magnetic fields strong enough to make a
permanent asymmetry ($B\ll 10^{10}\,{\rm G}$).

Type I X-ray bursts have provided a new way to determine the spin of
these neutron stars. Observations with the Proportional Counter Array
(PCA) on the {\it Rossi X-Ray Timing Explorer (RXTE)} of neutron stars
in six LMXBs have shown coherent oscillations during Type I X-ray
bursts, with frequencies $\nu_0$ that range from 300 to 600 Hz (see
Table \ref{tab:obs}). The simplest interpretation is that the burning
is not spherically symmetric, providing a temporary asymmetry on the
neutron star that allows for a direct measurement of rotation.
The coherent nature of the periodicities ($Q\gtrsim 300$),
large modulation amplitudes and stability of the frequency over at
least a year support its interpretation as the neutron star spin
(Strohmayer 1999b, and references therein).

It is expected theoretically that the burning should not be spherically
symmetric. Joss (1978) and Shara (1982) suggested that ignition of a
burst occurs at a local spot on the star, and not simultaneously over
the whole surface. This is because it takes hours to days to accumulate
the fuel, but only a few seconds for the thermal instability to grow.
Simultaneous ignition thus requires synchronization of the thermal state
of the accreted envelope to one part in $10^3$--$10^4$ over the
surface. More likely is that ignition is local and a burning front then
spreads laterally (Fryxell \& Woosley 1982; Nozakura, Ikeuchi, \&
Fujimoto 1984; Bildsten 1995), burning the rest of the accreted fuel and
creating a temporary brightness asymmetry on the neutron star surface
(Schoelkopf \& Kelley 1991; Bildsten 1995). Consistent with the picture
of a spreading burning front, the pulsation amplitude is observed to
decrease during the burst rise while the emitting area increases,
reaching a constant value during the decay (Strohmayer, Zhang \& Swank
1997b; Strohmayer et al. 1998c).

The oscillations during the burst rise are well-explained by the picture
of a spreading hotspot. However, several mysteries remain. First,
oscillations are often seen in the burst tail, when the whole surface of
the star has presumably ignited. The cause of the azimuthal asymmetry at
late times is not understood. Secondly, for those objects with
$\nu_0\approx 550\ {\rm Hz}$ in Table \ref{tab:obs}, there is evidence
that the burst oscillation frequency is twice the spin frequency of the
star, including observations of a $\approx 290\ {\rm Hz}$ subharmonic
(Miller 1999) and the fact that the kHz QPO separation is approximately
half the burst oscillation frequency in these sources (see van der Klis
2000 for a review). What might cause an $m=2$ azimuthal asymmetry is
unknown. Third, oscillations are not seen in all sources, or all Type I
bursts from the same source.

An initially puzzling feature of the observations was that the
oscillation frequency often changes during the burst, increasing by
$\Delta\nu\approx 1$--$2$ Hz in the burst tail. Strohmayer et
al. (1997a) proposed a simple explanation --- that this frequency
shift results from angular momentum conservation. The slight
hydrostatic radial expansion (contraction) of the burning layers as
the temperature increases (decreases) results in spin down (spin up)
if angular momentum is conserved. The time for radial heat transport
from the burning layers to the photosphere is about one second, which
means that the layers are puffed up and spun down by the time the
observer sees the burst. As the burning layers cool during the tail of
the burst, they contract and spin up. Strohmayer \& Markwardt (1999)
and Miller (1999, 2000) have modelled the observed frequency
evolution. They find that, in the evolving frame of the burning shell,
the oscillations are coherent, as expected if they are due to
rotation. Calculations show that the change in thickness of the
burning layers during a burst is $\Delta z\approx 20\,{\rm m}$ (Hanawa
\& Fujimoto 1984; Ayasli \& Joss 1982; Bildsten 1998). A simple
estimate of the spin down of the burning shell due to this change in
thickness is then $\Delta\nu\approx\nu_s (2\Delta z/R)\approx 1\,{\rm
Hz}\,(\nu_s/300\,{\rm Hz})(\Delta z/20\,{\rm m})(10\,{\rm km}/R)$
where $R$ is the neutron star radius and $\nu_s$ the spin
frequency. This roughly agrees with the observed values (Table
\ref{tab:obs}).

In this paper, we make a first attempt at understanding the evolution of
the neutron star atmosphere during a Type I X-ray burst on a rotating
neutron star. We use hydrostatic models of the neutron star atmosphere
to calculate the expansion and resulting spin evolution during a
burst. Our aim is to investigate whether the observed frequency changes
during bursts are consistent with spin down of the burning shell, thus
lending support to the interpretation that the burst oscillation
frequency is intimately related to the neutron star spin. This simple
picture demands that the hot burning material decouples from the cooler
underlying ashes and conserves its angular momentum, completing a few
phase wraps with the underlying star during the burst. We thus examine
mechanisms that might couple the burning layers to the neutron star, and
ask whether it is plausible that the burning layers can remain decoupled
for the $\approx 10\,{\rm s}$ duration of the burst.

We stress that we consider only one-dimensional models for these
hydrostatic and coupling calculations. We do not consider the complex
question of how the burning front spreads around the star during the
burst rise, what determines the number of hotspots on the surface, or
what causes the asymmetry at late times during the cooling tail of the
burst. We leave these questions for future investigations.

We start in \S 2 by describing the hydrostatic structure of the
atmosphere during fuel accumulation and the X-ray burst. In \S 3, we
calculate the expansion of the atmosphere and show that the expected
spin down of a decoupled layer is consistent with observed values. We
consider heat transport through the atmosphere, and ask how a single
coherent frequency is transmitted to the observer. In \S 4, we discuss
hydrodynamic mechanisms that could transport angular momentum within the
burning layers or couple the hot burning material to the underlying
colder and denser ashes. In \S 5, we summarize our results and discuss
the many remaining puzzles. Finally, we present our conclusions in \S 6.

\section{Thermal Structure and Expansion of the Burning Layers}

Most neutron stars in LMXBs accrete hydrogen and helium rich material
from their companions at rates $\dot M\sim 10^{-11}-10^{-8} M_\odot \
{\rm yr^{-1}}$. For accretion rates $\dot M\gtrsim 2\times
10^{-10}\,M_\odot \ {\rm yr^{-1}}$ (see Bildsten 1998 and references
therein), the accumulating hydrogen is thermally stable and burns via
the hot CNO cycle of Hoyle \& Fowler (1965). The temperature of most of
the atmosphere is $\gtrsim 8\times 10^7\,{\rm K}$ so that the time for a
proton capture onto a $^{14}$N nucleus is less than the time for the
subsequent beta-decays. This fixes the energy production rate at the
value
\begin{equation} \label{eq:CNO} \epsilon_H=5.8\times
10^{13} \left({\zcno\over 0.01}\right) {\rm \ ergs \ g^{-1} \ s^{-1}},
\end{equation} 
where $\zcno$ is the mass fraction of CNO nuclei. This energy
production rate is independent of temperature or density, and simply
set by the beta-decay timescales of $^{14}$O (half-life 71 s) and
$^{15}$O (half-life 122 s). Because the hydrogen burns at a constant
rate, the time to burn all of it depends only on the metallicity and
initial hydrogen abundance. The hydrogen mass fraction $X$ in a given
fluid element changes at a rate $dX/dt=-\epsilon_H/E_H$, where
$E_H\approx 6.7 \,{\rm MeV}/{\rm proton}\approx 6.4\times
10^{18}\,{\rm erg\,g^{-1}}$ is the energy release per gram from
burning hydrogen to helium. The time to burn all the hydrogen is then
\begin{equation}
t_H\approx 22\ {\rm hours}\ \left({0.01\over\zcno}\right)
\left({X_0\over 0.71}\right),
\end{equation}
where $X_0$ is the initial hydrogen mass fraction.

The X-ray burst is triggered when helium burning becomes unstable at
the base of the accumulated layer, at a density $\sim
10^5$--$10^6\,{\rm g\,cm^{-3}}$ and temperature $\approx 2\times
10^8\,{\rm K}$. The composition at the base of the layer depends on
how much hydrogen has burned during the accumulation (Fujimoto, Hanawa
\& Miyaji 1981, hereafter FHM; Bildsten 1998), which is determined by
the local accretion rate, $\dot m$. The local Eddington accretion rate
is $\dot m_{\rm Edd}=2m_pc/(1+X)R\sigma_{\rm Th}$, where $\sigma_{\rm
Th}$ is the Thomson scattering cross-section, $m_p$ is the proton
mass, $c$ is the speed of light, and $R$ is the stellar radius. In
this paper, we use the Eddington accretion rate for solar composition
($X_0=0.71$) and $R=10$~km, $\dot m_{\rm Edd}\equiv 8.8\times 10^4\
{\rm g \ cm^{-2} \ s^{-1}}$, as our basic unit for the local accretion
rate (for a 10 km neutron star, this corresponds to a global rate
$\dot M=1.7\times 10^{-8}\,M_\odot\,{\rm yr^{-1}}$). For
$\dot{m}\gtrsim 0.03\,\dot{m}_{\rm Edd}$, the helium burning becomes
unstable before all the hydrogen is burned and the helium ignites and
burns in a hydrogen rich environment. At lower accretion rates
$\dot{m}\lesssim 0.03\,\dot{m}_{\rm Edd}$, there is enough time to
burn all the hydrogen, and a pure helium layer accumulates which
eventually ignites.  We now describe simple models of the atmosphere
immediately prior to (\S 2.1) and during (\S 2.2) the X-ray burst for
both of these cases. At lower accretion rates still, $\dot{m}\lesssim
0.01\dot{m}_{\rm Edd}$, the hydrogen burning becomes unstable and
triggers a mixed hydrogen/helium burning flash (FHM).

\subsection{The Accumulating Atmosphere}

\begin{deluxetable}{llccllclc}
\small
\tablecolumns{9}
\tablecaption{Plane Parallel Ignition Conditions\tablenotemark{a}\label{tab:ign}}
\tablewidth{0pt}
\tablehead{
\colhead{$\dot{m}/\dot{m}\edd$\tablenotemark{b}} 
& \colhead{$Z_{\rm CNO}$} &
\colhead{$T$} & \colhead{$y$} & 
\colhead{$X$} & \colhead{$Y$} & \colhead{$\rho$} &
\colhead{$y/\dot{m}$\tablenotemark{c}} & \colhead{$\Delta z$(90\%)\tablenotemark{d}}\nl
\colhead{} & \colhead{} & \colhead{($10^8\,{\rm K}$)} &
\colhead{($10^8\,{\rm g\,cm^{-2}}$)} & \colhead{} & \colhead{} &
\colhead{($10^6\,{\rm g\,cm^{-3}}$)} & \colhead{(h)} &
\colhead{(m)}}
\startdata
\multicolumn{9}{c}{Pure He Ignition}\nl
\hline\nl
0.01  & 0.005 & 1.49 & 5.40 & 0.0  & 0.995 & 2.28 & 170 & 4.2\\
      & 0.01  & 1.35 & 12.9 & 0.0  & 0.99  & 4.14 & 407 & 5.1\\
      & 0.02  & 1.22 & 34.0 & 0.0  & 0.98  & 8.01 & 1073 & 6.9\\
0.015 & 0.01  & 1.56 & 3.67 & 0.0  & 0.99 & 1.75 & 77 & 3.8 \\
      & 0.02  & 1.41 & 8.50 & 0.0  & 0.98 & 3.11 & 179 & 4.4 \\
0.02  & 0.01  & 1.74 & 1.92 & 0.0  & 0.99 & 1.12 & 30 & 3.8 \\
      & 0.02  & 1.57 & 3.62 & 0.0  & 0.98 & 1.74 & 57 & 3.5  \\
0.03  & 0.02  & 1.83 & 1.56 & 0.0  & 0.98 & 0.97 & 16 & 3.6  \\

\cutinhead{Mixed H/He Ignition}
0.015 & 0.005 & 1.73 & 2.05 & 0.01 & 0.99 & 1.16 & 43 & 4.2\\
0.02  & 0.005 & 1.78 & 2.18 & 0.15 & 0.85 & 1.04 & 34 & 4.5\\
0.03  & 0.005 & 1.86 & 2.35 & 0.31 & 0.69 & 0.94 & 25 & 4.9\\
      & 0.01  & 1.89 & 1.67 & 0.14 & 0.85 & 0.87 & 18 & 4.2\\
0.1   & 0.005 & 2.05 & 2.70 & 0.57 & 0.42 & 0.84 & 8.5 & 5.6 \\
      & 0.01  & 2.13 & 2.04 & 0.50 & 0.49 & 0.71 & 6.4 & 5.1\\
      & 0.02  & 2.20 & 1.51 & 0.40 & 0.58 & 0.62 & 4.8 & 4.7\\
0.3   & 0.005 & 2.24 & 2.62 & 0.67 & 0.33 & 0.76 & 2.8 & 5.8\\
      & 0.01  & 2.33 & 2.14 & 0.64 & 0.35 & 0.66 & 2.3 & 5.5\\
      & 0.02  & 2.44 & 1.70 & 0.59 & 0.39 & 0.57 & 1.8 & 5.2\\
\enddata
\tablenotetext{a}{Conditions at the base of the accumulated column at
ignition. We take $M=1.4\,M_\odot$, $R=10\,{\rm km}$, and
$g=GM/R^2=1.9\times 10^{14}\,{\rm cm\,s^{-2}}$. The ignition pressure
is $P=gy=1.9\,y_8\times 10^{22}\,{\rm erg\,cm^{-3}}$ and the
accumulated mass over the whole surface is $\Delta M=4\pi R^2y=1.3\,y_8\times 10^{21}\,{\rm
g}$. We take the flux from deeper in the star to be $F_b=150$ keV per
nucleon.}
\tablenotetext{b}{$\dot m\edd=8.8\times 10^4\,{\rm
g\,cm^{-2}\,s^{-1}}$, equivalent to a global rate $\dot
M\edd=1.7\times 10^{-8}\,M_\odot\,{\rm yr^{-1}}$.}
\tablenotetext{c}{The time to accumulate the unstable column.}
\tablenotetext{d}{The height above the base which contains 90\% of the
mass.}
\end{deluxetable}

The neutron star atmosphere is in hydrostatic balance as the accreted
hydrogen and helium accumulates. The pressure obeys $dP/dz=-\rho g$,
where $\rho$ is the density and the gravitational acceleration
$g\equiv GM/R^2$ is constant in the thin envelope. A useful variable
is the column depth $y$ (units g cm$^{-2}$), defined by $dy\equiv
-\rho dz$, giving $P=gy$. As the matter accumulates, a given fluid
element moves to greater and greater column depth. In this paper, we
take $g_{14}\equiv g/10^{14}\ {\rm cm\,s^{-2}}=1.9$, appropriate for a
$M=1.4\,M_\odot$ and $R=10\,{\rm km}$ neutron star. As described
above, the hydrogen burning rate is a constant so that the change of
$X$ with column depth $y$ is $dX/dy=-\epsilon_H/\dot{m}E_H$, where we
take $E_H\equiv 6.4\times 10^{18}\ {\rm erg\ g^{-1}}$ (we have
neglected the $\approx 0.5\ {\rm MeV/proton}$ lost as neutrinos in the
hot CNO cycle, see Wallace \& Woosley 1981). Integrating this
equation, we find the hydrogen abundance as a function of depth is
\begin{equation}\label{eq:X}
X(y)=\cases{ X_0\left[1-(y/y_d)\right] & $y>y_d$, \cr 
0 & $y<y_d$, \cr}
\end{equation}
where $X_0$ is the initial hydrogen abundance, and the column depth at
which the hydrogen runs out is $y_d\equiv X_0\dot{m}E_H/\epsilon_H$,
or
\begin{equation}\label{eq:yd}
y_d=6.8\times 10^8\,{\rm g\,cm^{-2}}
\left({\dot{m}\over 0.1\dot{m}\edd}\right) \left({0.01\over Z_{\rm
CNO}}\right)
\left({X_0\over 0.71}\right).
\end{equation}
If helium ignites at a column depth $y<y_d$, a mixed hydrogen/helium
burning flash occurs, otherwise a pure helium layer accumulates which
eventually ignites at a column $y>y_d$.

The thermal profile of the accumulating layer is described by the heat
equation,
\begin{equation}\label{eq:heat}
{dT\over dy}={3\kappa F\over 4acT^3},
\end{equation}
where $\kappa$ is the opacity and $F$ is the outward heat flux. The
entropy equation is
\begin{equation}
\epsilon+{dF\over dy}=c_P{\partial T\over\partial t}+
{c_PT\dot{m}\over y} \left({d\ln T\over d\ln y}-n\right)
\end{equation}
(Bildsten \& Brown 1997), where $n\equiv (d\ln T/d\ln P)_S$, $c_P$ is
the heat capacity at constant pressure, and the terms on the right
hand side describe the compression of the accumulating matter. The
compressional terms contribute $\approx c_PT\approx 5k_BT/2\mu m_p\approx
20\,T_8\ {\rm keV}$ per accreted nucleon, where $\mu$ is the mean
molecular weight, and $T_8\equiv T/10^8\ {\rm K}$. In comparison, the
hot CNO cycle hydrogen burning gives $\approx 7 X_0\ {\rm MeV}$ per
accreted nucleon for pure helium ignition, or $\approx 7 X_0(y_b/y_d)\
{\rm MeV}$ for mixed H/He ignition, where $y_b$ is the column depth at
the base. There is additional flux from heat released by electron
captures and pycnonuclear reactions deep in the crust, giving $\approx
100\ {\rm keV}$ per nucleon (Brown \& Bildsten 1998; Brown 2000). For
the purposes of this paper, we neglect the compressional terms in the
entropy equation, and take
\begin{equation}\label{eq:entropy}
\epsilon=-{dF\over dy},
\end{equation}
but include a flux at the base ($y=y_b$) of $F_b=150\ {\rm keV}$ per
nucleon as an approximation to the heat from the crust and
compressional heating.

We find the temperature profile by integrating equations~(\ref{eq:heat})
and (\ref{eq:entropy}) from the top of the atmosphere to the base at
column depth $y_b$. At the top, which we arbitrarily place at $y=5\times
10^4\,{\rm g\,cm^{-2}}$, we set the temperature using the analytic
radiative zero solution for a constant flux atmosphere with Thomson
scattering opacity (the solutions are not sensitive to this upper
boundary condition).  For $y<y_d$ we take $\epsilon=\epsilon_H$, and for
$y>y_d$ we set $\epsilon=0$. The flux at the top $F_t$ is set by the
energy release from hot CNO burning in the atmosphere and the flux at
the base, giving $F_t=F_b+\epsilon_Hy_b$ or $F_t=F_b+\epsilon_Hy_d$,
whichever is smaller\footnote{As pointed out by Taam \& Picklum (1978),
some helium burning occurs during accumulation, which increases the
number of CNO nuclei at ignition (for example, see Fujimoto et al. 1987,
Figure 4). FHM and Hanawa \& Fujimoto (1982) estimated the increase in
CNO abundance, concluding that the effect on the ignition temperature
and density would be small because of the large temperature sensitivity
of the triple alpha reaction. In calculations which include helium
burning during accumulation, we find that helium burning reactions
decrease the ignition column by 10--20\%, increase the ignition
temperature by less than a few percent and increase the metallicity at
ignition by factors of 2--3. For the purposes of this paper, we
adopt simple models with hot CNO burning only.}. The opacity $\kappa$
has contributions from electron scattering, free-free absorption and
electron conduction, which we calculate as described in Schatz et
al. (1999). Since in the hot CNO cycle the seed nuclei are mostly
$^{14}$O or $^{15}$O waiting to $\beta$-decay, we take the gas to be a
mixture of hydrogen (mass fraction $X$ given by eq.[\ref{eq:X}]),
$^{14}$O and $^{15}$O (mass fraction $Z_{\rm CNO}$) and helium (mass
fraction $Y=1-X-Z_{\rm CNO}$). The ratio by number of $^{14}$O to
$^{15}$O is given by the ratio of the beta-decay timescales, giving
$Z_{14}=0.352 Z_{\rm CNO}$ and $Z_{15}=0.648 Z_{\rm CNO}$. To obtain a
simple analytic estimate of the base temperature, we first integrate
equation (\ref{eq:entropy}) with $\epsilon_H$ a constant to find
$F(y)=\epsilon_H(y_b-y)$, where we ignore compressional heating and the
flux from the base. Substituting this into equation (\ref{eq:heat}) and
integrating assuming constant opacity, we find
$T^4(y)=(3\kappa/ac)\epsilon_Hyy_b(1-y/2y_b)$, giving
\begin{eqnarray}
T_b\approx 2.6\times 10^8\,{\rm K}\, \left({\kappa\over
0.08\ {\rm cm^2 \ g^{-1}}}\right)^{1/4} \left({Z_{\rm CNO}\over 0.01}\right)^{1/4}\nonumber\\
\left({y_b\over 2\times 10^8\,{\rm g\,cm^{-2}}}\right)^{1/2},
\end{eqnarray}
where we have inserted a typical value for $\kappa$.

\begin{figure}
\plottwo{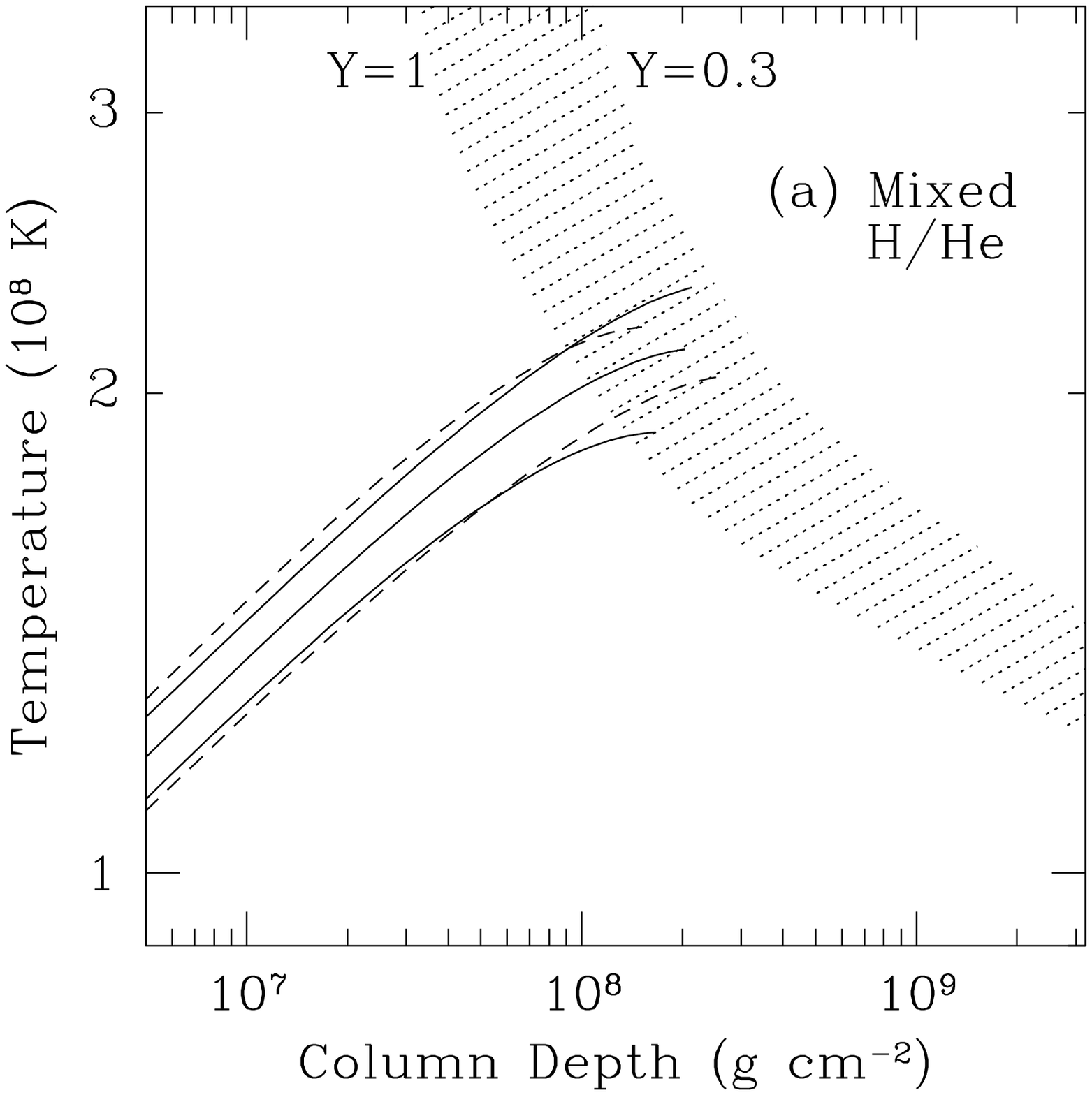}{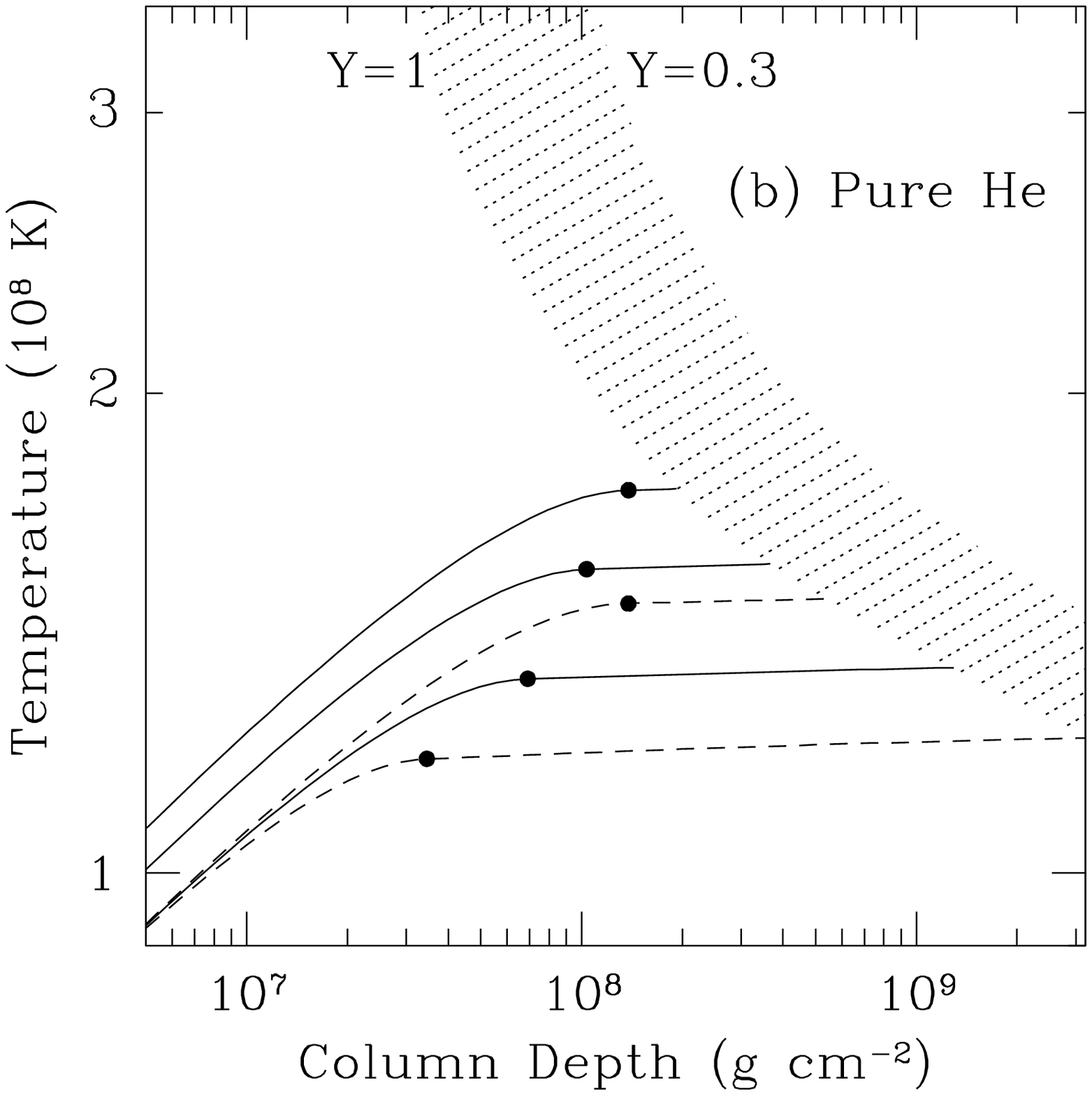}
\caption{Profiles of the atmosphere immediately prior to (a) mixed
hydrogen and helium ignition and (b) pure helium ignition. The hatched
region shows where helium ignition occurs for helium mass fraction $Y$
between 1.0 and 0.3. All models have a base flux of $F_b=150\,{\rm
keV}$ per nucleon (see text). (a) The solid lines show the profiles
for $Z_{\rm CNO}=0.01$ and (bottom to top) $\dot m/\dot
m\edd=0.03,0.1,0.3$. The dashed lines show the $\dot m=0.1\,\dot
m\edd$ case for $Z_{\rm CNO}=0.005$ (bottom curve) and $0.02$ (top
curve). (b) The solid lines show the profiles for $Z_{\rm CNO}=0.01$
and (bottom to top) $\dot m/\dot m\edd=0.01,0.015,0.02$. The dashed
lines show the $\dot m=0.01\,\dot m\edd$ case for $Z_{\rm CNO}=0.005$
(top curve) and $0.02$ (bottom curve). The black dots show where the
hydrogen runs out (eq. [\ref{eq:yd}]).
\label{fig:settle}}
\end{figure}

We find the column depth at the base of the accumulated layer just
before the thermally unstable helium ignition by comparing the
temperature sensitivity of the heating and cooling rates (FHM; Fushiki
\& Lamb 1987b; Bildsten 1998). The heating rate due to the
triple-alpha reaction is
\begin{equation}
\epsilon_{3\alpha}=5.3\times 10^{21}\,{\rm erg\,g^{-1}\,s^{-1}}\ 
f{\rho_5^2Y^3\over T_8^3}\exp\left({-44\over T_8}\right),
\end{equation}
where $f$ is the screening factor (Fushiki \& Lamb 1987a).  In
addition, $^{12}$C rapidly captures protons once it is made, increasing
the energy release from the triple alpha reaction. To account
for this, we multiply $\epsilon_{3\alpha}$ by a factor
$1+Q_{12}/Q_{3\alpha}=1.9$, where $Q_{3\alpha}=7.274$ MeV,
$Q_{12}=Q(^{12}{\rm C}+2p\rightarrow^{14}{\rm O})=6.57$ MeV. An
effective local cooling rate is obtained from equations
(\ref{eq:heat}) and (\ref{eq:entropy}), giving
\begin{equation}
\epsilon_{\rm cool}\approx {acT^4\over 3\kappa y^2}.
\end{equation}
As the matter accumulates, the column depth at the base $y_b$
increases until $d\epsilon_{3\alpha}/dT>d\epsilon_{\rm cool}/dT$, at
which point a thermal instability occurs (FHM). We find the
temperature profile at ignition by choosing $y_b$ such that
$d\epsilon_{3\alpha}/dT=d\epsilon_{\rm cool}/dT$ at the base. In
Figure \ref{fig:settle}, the hatched region in the temperature-column
depth plane indicates where helium ignition occurs for abundances
ranging from the initial value ($Y=0.3$) to pure helium ($Y=1.0$).

The conditions at the base of the accumulated column at the time of
ignition for a variety of metallicities and accretion rates are shown
in Table \ref{tab:ign}. We have separated our solutions into two
groups depending on whether the helium unstably ignites before (mixed
H/He ignition) or after (pure He ignition) the hydrogen is completely
burned. We give the temperature, density and composition at the base,
the time to accumulate the unstable column $y_b/\dot m$, and the
physical distance from the base to the place where $y=y_b/10$, $\Delta
z(90$\%$)$. For mixed H/He (pure He) ignition, $\Delta
z(90$\%$)\approx 5\ {\rm m}$ ($3.5$--$7\ {\rm m}$).

Figure \ref{fig:settle}(a) shows the temperature profile of models
with mixed H/He ignition. The solid lines are for $Z_{\rm CNO}=0.01$
and (bottom to top) $\dot m/\dot m\edd=0.03,0.1,0.3$. The dashed lines
are for $\dot m=0.1\,\dot m\edd$ and $Z_{\rm CNO}=0.005$ (bottom
curve) or $0.02$ (top curve).  Figure \ref{fig:settle}(b) shows
temperature profiles for pure He ignition. The solid lines show
$Z_{\rm CNO}=0.01$ and (bottom to top) $\dot m/\dot m\edd=0.01,0.015,$
and $0.02$. The dashed lines are for $\dot m=0.01\,\dot m\edd$ and
$Z_{\rm CNO}=0.005$ (top curve) or $0.02$ (bottom curve).  The black
dots show the depth where hydrogen runs out (at a column $y_d$,
eq. [\ref{eq:yd}]).

For the mixed H/He ignition models, the ignition temperature and density
do not depend sensitively on $\dot m$. Slight differences arise because
at higher accretion rates less helium is made from hydrogen burning,
requiring a higher density and temperature for ignition.  For pure He
ignition, the ignition column and temperature are much more sensitive to
accretion rate. After the hydrogen is burned, there is a slight
temperature gradient to carry the flux from the base $F_b$, but the
temperature at ignition is mainly set by the temperature at the base of
the hydrogen burning shell (FHM; Wallace, Woosley and Weaver 1982). This
temperature is greater at higher accretion rates, leading to a smaller
column depth at ignition.

\begin{deluxetable}{lllcccccll}
\small
\tablecolumns{10}
\tablecaption{Models with a Convection Zone\label{tab:conv}}
\tablewidth{0pt}
\tablehead{
\colhead{$(y_b-y_c)/y_b$\tablenotemark{a}} & \colhead{$T_b$} &
\colhead{$F/F\edd$\tablenotemark{c}} & \colhead{$\Delta z_c$} &
\colhead{$\Delta z$(90\%)} & \colhead{$\rho_b$} &
\colhead{$\beta_b$\tablenotemark{b}} &
\colhead{$\bar{X}$} & \colhead{$\bar{Y}$} & \colhead{$\mu$} \\
\colhead{} & \colhead{($10^9\,{\rm K}$)} & \colhead{} & \colhead{(m)} &
\colhead{(m)} & \colhead{($10^5\ {\rm g\ cm^{-3}}$)} & 
\colhead{} & \colhead{} & \colhead{} & \colhead{}}
\startdata
\multicolumn{10}{c}{Mixed Ignition} \nl
\multicolumn{10}{c}{($\dot m=0.1\ \dot m\edd$; $Z_{\rm
CNO}=0.01$; $y_b=2.04\times 10^8\ {\rm g\,cm^{-2}}$; $\Delta
z(90$\%$)=5.1\,{\rm m}$)} \nl
\hline\nl
0.99 & 1.7 & 0.89 & 61.5 & 40.9 & 0.83 & 0.46 & 0.60 & 0.39 & 0.67 \\
0.95 & 1.7 & 1.15 & 48.5 & 40.7 & 0.84 & 0.46 & 0.60 & 0.39 & 0.67 \\
0.8  & 1.5 & 0.74 & 18.3 & 24.4 & 1.41 & 0.67 & 0.58 & 0.41 & 0.68 \\
0.5  & 1.2 & 0.31 & 5.5  & 15.9 & 2.30 & 0.86 & 0.55 & 0.44 & 0.70 \\
\hline\nl
\multicolumn{10}{c}{Pure He Ignition}\nl
\multicolumn{10}{c}{($\dot m=0.015\ \dot m\edd$; $Z_{\rm
CNO}=0.01$; $y_b=3.67\times 10^8\ {\rm g\,cm^{-2}}$; $\Delta
z(90$\%$)=3.8\,{\rm m}$)} \nl
\hline\nl
0.99 & 2.0  & 1.06 & 45.6 & 30.2 & 2.04 & 0.42 & 0.09 & 0.90 & 1.16 \\
0.95 & 1.95 & 1.21 & 29.8 & 25.0 & 2.45 & 0.45 & 0.07 & 0.92 &1.20 \\
0.8  & 1.6 & 0.59 & 8.9 & 13.1 & 5.16 & 0.76 & 0.01 & 0.98 &1.32 \\
0.5  & 1.3 & 0.35 & 3.1 & 10.0 & 7.38 & 0.90 & 0.00 & 0.99 & 1.35 \\
\enddata
\tablenotetext{a}{The fraction of the accumulated mass that
becomes convective.}
\tablenotetext{b}{The ratio of gas pressure to total pressure at the
base.}
\tablenotetext{c}{$F\edd=1.67\times 10^{25}\ {\rm erg\ cm^{-2}\ s^{-1}}$.}
\end{deluxetable}

The effect of increasing metallicity for mixed H/He ignition models is
to increase the energy generation rate, and thus the temperature at the
base. Together with the increased rate of helium production, this allows
He ignition at a lower column depth, as seen in Figure
\ref{fig:settle}(a). For pure He ignition, the effect of increasing
metallicity is exactly opposite. The ignition temperature is lowered
with increasing metallicity because the hydrogen runs out at a smaller
column, giving a smaller temperature at the base of the hydrogen burning
shell. Ignition then requires a greater column depth at the base of the
almost isothermal pure helium layer, as seen in Figure
\ref{fig:settle}(b).

The $\dot m$ at which the transition between pure He and mixed H/He
ignitions occurs depends on metallicity. The $\dot m=0.02\ \dot m\edd$
case in Figure \ref{fig:settle} (right panel) just burns the hydrogen
before ignition; this is the transition $\dot m$ for a metallicity
$Z_{\rm CNO}=0.01$. For $Z_{\rm CNO}=0.005$, the transition occurs for
$\dot m\approx 0.01\ \dot m\edd$, and $\dot m\gtrsim 0.03\ \dot m\edd$
for $Z_{\rm CNO}=0.02$. This is in rough agreement with the $\propto
Z_{\rm CNO}^{13/18}$ scaling estimated by Bildsten (1998). The
transition accretion rate increases with increasing metallicity because
the time to burn all the hydrogen decreases, allowing a pure helium
layer to build up even at large accretion rates.

Our results agree well with those of previous authors. The
calculations we have presented here are similar to those of Taam
(1980) and Hanawa \& Fujimoto (1982). Hanawa \& Sugimoto (1982)
simulated pure He ignition bursts on neutron stars with $g_{14}=0.93$
and $7.1$. Conditions at the base of the hydrogen burning shell that
they find agree with our calculations to 10\%, and the helium ignition
column agrees to within a factor of two. We find similar agreement
with the pure He ignition calculations of Wallace, Woosley \& Weaver
(1982, hereafter W82). However, the density at the base of the
hydrogen burning shell that we compute is twice as great as that given
in Table 1 of W82, whereas the pressure and temperature agree. We do
not know the reason for this discrepancy, but it may be that W82 used
a fixed solar composition for the hydrogen burning shell, giving a
different value for $\mu_e$ at the base. Hanawa \& Fujimoto (1984)
simulated mixed H/He ignition bursts. Taking into account their
gravity, $g_{14}=3.3$, our ignition column and temperature agree with
their calculations to 10--30\%.

\subsection{The Atmosphere During the Burst}\label{sec:burstmodels}

Many authors have simulated X-ray bursts for both mixed H/He and pure He
ignitions, for reviews see Lewin, van Paradijs, \& Taam (1993), and
Bildsten (1998). In this section, we make simplified models which allow
us to calculate the hydrostatic expansion of the atmosphere during a
burst. We do not discuss the lateral spreading of the burning front in
this paper. In addition, our models are not appropriate for the radius
expansion phase of bursts with super-Eddington luminosities (radius
expansion bursts; Lewin, van Paradijs, \& Taam 1993). We discuss the
effect of radius expansion on the observability of oscillations in \S
\ref{sec:radiusexp}. We start in \S \ref{sec:convmodels} by considering
the convective stage of a burst. As we discuss later, convection is
important because it likely enforces rigid rotation, and may affect the
observability of a coherent signal (\S \ref{sec:expansion}). In \S
\ref{sec:radmodels} we compute fully-radiative atmospheric models,
appropriate for those bursts that do not become convective, or for the
later stages of a convective burst, when the convection zone retreats.

\subsubsection{Models with Convection}\label{sec:convmodels}

\begin{deluxetable}{lccccc}
\small
\tablecolumns{6}
\tablecaption{Fully Radiative Models\label{tab:noconv}}
\tablewidth{0pt}
\tablehead{
\colhead{$F/F\edd$} &\colhead{$T_b$} & \colhead{$\Delta z$(90\%)} & 
\colhead{$\rho_b$} & \colhead{$\beta_b$} & \colhead{Notes} \\
\colhead{} & \colhead{($10^9\,{\rm K}$)} & \colhead{(m)} & 
\colhead{($10^5\ {\rm g\ cm^{-3}}$)} & \colhead{}
& \colhead{}}
\startdata
\multicolumn{6}{c}{Mixed Ignition} \nl
\multicolumn{6}{c}{($\dot m=0.1\ \dot m\edd$; $Z_{\rm
CNO}=0.01$; $y_b=2.04\times 10^8\ {\rm g\,cm^{-2}}$; $\Delta
z(90$\%$)=5.1\,{\rm m}$)} \nl
\hline\nl
1.15 & 1.49 & 29.7 & 1.53 & 0.68 & a\\
     & 1.29 & 25.7 & 2.13 & 0.82 & b\\
0.74 & 1.35 & 22.3 & 1.94 & 0.78 & a\\
     & 1.17 & 20.1 & 2.50 & 0.88 & b\\
0.31 & 1.12 & 15.6 & 2.67 & 0.90 & a\\
     & 0.96 & 14.5 & 3.20 & 0.94 & b\\
\hline\nl
\multicolumn{6}{c}{Pure He Ignition}\nl
\multicolumn{6}{c}{($\dot m=0.015\ \dot m\edd$; $Z_{\rm
CNO}=0.01$; $y_b=3.67\times 10^8\ {\rm g\,cm^{-2}}$; $\Delta
z(90$\%$)=3.8\,{\rm m}$)} \nl
\hline\nl
1.21 & 1.57 & 16.9 & 5.44 & 0.78 & a\\
     & 1.38 & 15.4 & 6.83 & 0.87 & b\\
0.59 & 1.34 & 11.9 & 7.10 & 0.88 & a\\
     & 1.17 & 11.1 & 8.35 & 0.93 & b\\
0.35 & 1.19 & 9.9  & 8.22 & 0.93 & a\\
     & 1.04 & 9.4  & 9.39 & 0.96 & b\\
\enddata
\tablenotetext{a}{Constant flux.}
\tablenotetext{b}{Constant $\epsilon$.}
\end{deluxetable}

In one-dimensional models, the energy release from the temperature
sensitive helium burning reaction makes the early stages of many
bursts convective. The convection zone expands outwards, extending
over a few scale heights to the radiative outermost layers (Joss
1977).  For example, Hanawa \& Fujimoto (1984) computed a mixed H/He
ignition model, in which the atmosphere was convective for $\approx 1\
{\rm s}$ after ignition, reaching a maximum extent of $\approx 99\%$
of the accumulated mass. Hanawa \& Sugimoto (1982) found for pure He
ignition that a convection zone rapidly grew to encompass most of the
atmosphere, and then shrunk back, disappearing $\approx 1\ {\rm s}$
later when the nuclear fuel was almost exhausted.

The convection in the neutron star atmosphere is very efficient, since
the time for sound to cross a scale height ($c_s/g\sim 10^{-6}\,{\rm
s}$) is much shorter than the local thermal time ($\sim 1$--$10\,{\rm
s}$). Thus the atmosphere has a nearly adiabatic profile when
convective. For a given temperature $T_b$ and column depth $y_b$ at
the base, the thermal profile of the convective zone just follows the
adiabat $T=T_b(y/y_b)^n$, where $n\equiv (\partial\ln T/\partial\ln
P)_{S}$. We take $y_b$ to be the value given by the settling solution
at ignition; the convection zone profile is then determined by the
single parameter $T_b$ when the mean molecular weight, $\mu$, is
fixed. 

During the burst, the temperature at the base reaches $\gtrsim
10^9\,{\rm K}$, high enough that radiation pressure is important (Joss
1977).  At these temperatures, the degeneracy is lifted (we find the
electron pressure differs from the ideal gas value by less than a few
percent) so the total pressure at the base is
\begin{equation}
P_b={\rho_bk_BT_b\over\mu m_p}+{1\over 3}aT_b^4,
\end{equation}
where $\rho_b$ is the density
at the base. However, the total pressure is also fixed by the weight
of the overlying atmosphere, which gives $P_b=gy_b$. Thus $T_b$ cannot
exceed the critical value $T_{\rm max}=(3gy_b/a)^{1/4}$, or
\begin{equation}\label{eq:Tmax}
T_{\rm max}=2.2\times 10^9\,{\rm K}\,
\left({y_b\over 3\times 10^8\,{\rm g\,cm^{-2}}}\right)^{1/4}
\left({g_{14}\over 1.9}\right)^{1/4}.
\end{equation}
As $T_b$ approaches $T_{\rm max}$, radiation pressure becomes
increasingly important, forcing the density to decrease. We show below
that this greatly enhances the expansion of the atmosphere.

As clearly argued by Joss (1977), the convection zone cannot extend
all the way to the photosphere. A radiative layer is always needed to 
transport 
the heat to the photosphere. We assume that most of the burning occurs
in the convective layer, in which case the radiative atmosphere
carries a constant flux and is described by the heat equation
(\ref{eq:heat}) with constant $F$. For the convective zone, we
integrate $dT/dy=n(T/y)$ for a given base temperature $T_b$ (the
super-adiabaticity needed to carry the flux in the convective zone is
negligible when the convection is so efficient; see Cox \& Giuli
1968). For these simple models, we treat the flux $F$ and base
temperature $T_b$ as free parameters (in reality, the temperature
throughout the convective region determines the nuclear energy
generation rate and thus the flux). To compute the thermal profile of
the atmosphere, we integrate the convection zone outwards from the
base, and the radiative zone inwards from the surface, varying the
column depth of the top of the convection zone $y_c$ until the
temperature matches at the interface.

\begin{figure}
\plottwo{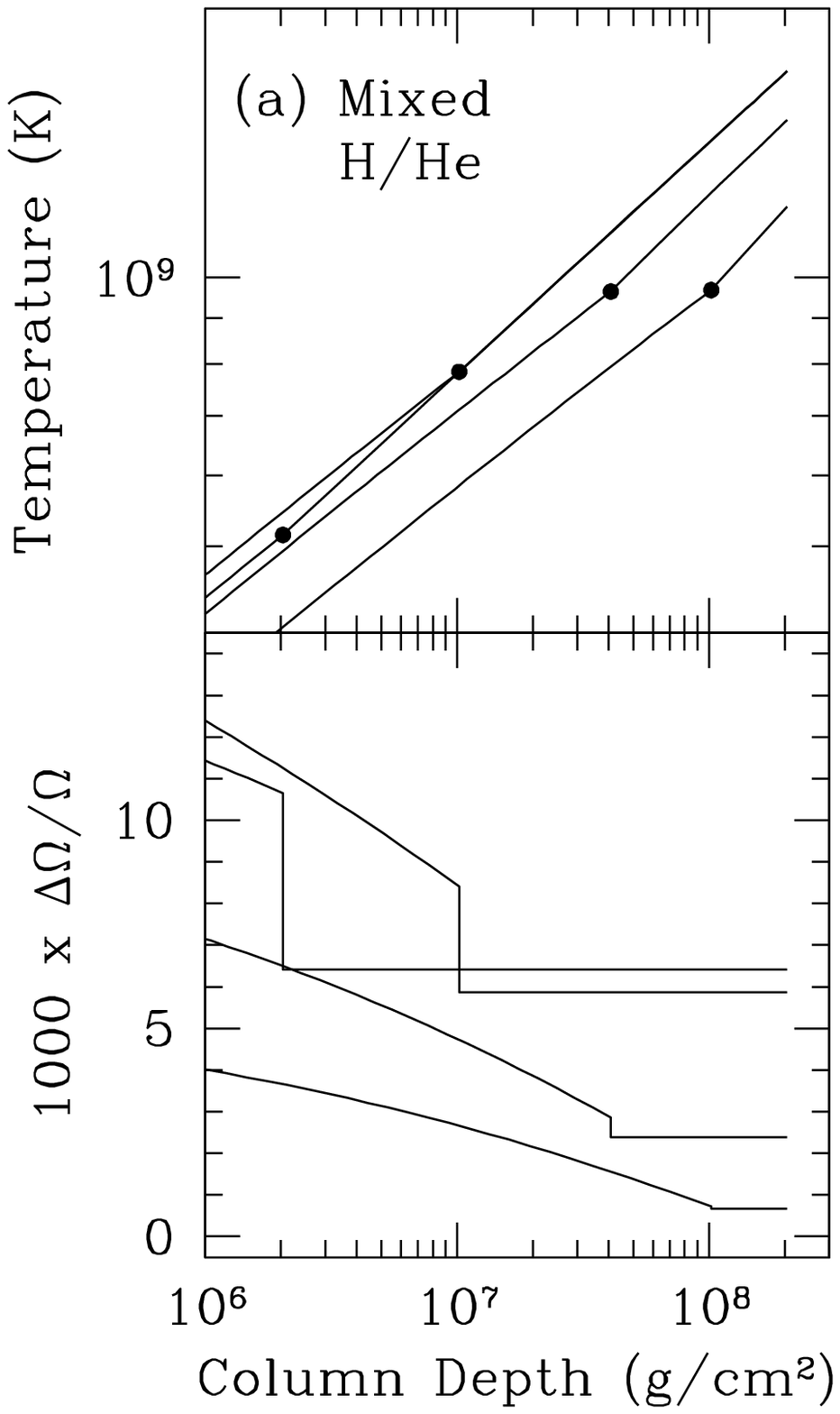}{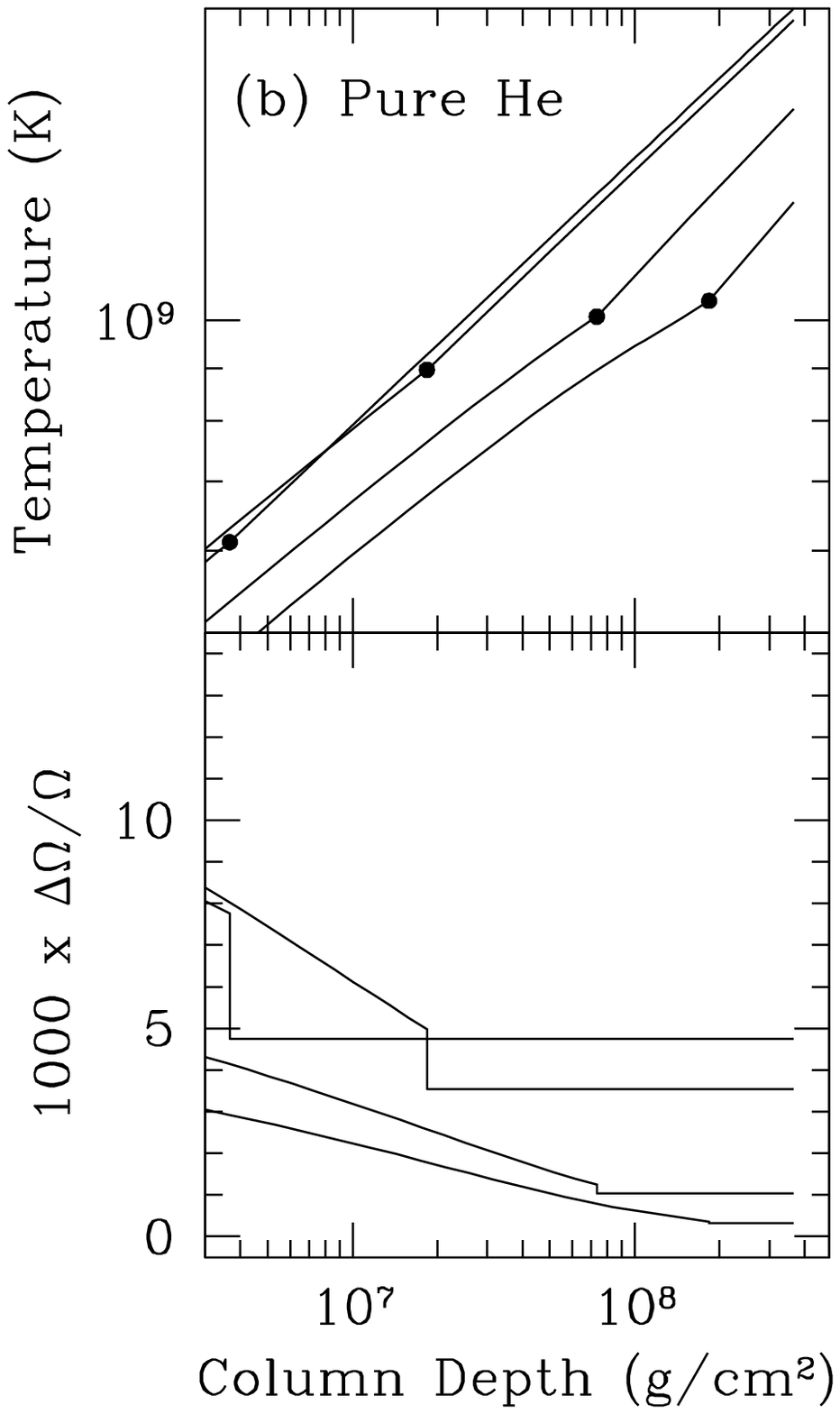}
\caption{Temperature profiles (upper panel) and spin down (lower
panel) for the convective models of Table \ref{tab:conv}. In each
case, the black dots mark the top of the convection zone at column
depth $y_c$ (Table \ref{tab:conv}). (a) We show models for mixed H/He
ignition with $\dot m=0.1\,\dot m\edd$, $Z_{\rm CNO}=0.01$, and
$y_b=2.04\times 10^8\,{\rm g\,cm^{-2}}$. These models have
$(y_b-y_c)/y_b=0.5,0.8,0.95$ and $0.99$, $T_b/10^9\,{\rm K}=1.2, 1.5,
1.7$ and $1.7$, and $F/F\edd=0.31, 0.74, 1.15$ and $0.89$. (b) We show
models for pure He ignition with $\dot m=0.015\,\dot m\edd$, $Z_{\rm
CNO}=0.01$ and $y_b=2.53\times 10^8\,{\rm g\,cm^{-2}}$. These models
have $(y_b-y_c)/y_b=0.5,0.8,0.95$ and $0.99$, $T_b/10^9\,{\rm K}=1.3,
1.6, 1.95$ and $2.0$, and $F/F\edd=0.35, 0.59, 1.21$ and $1.06$. The
difference in slope of the convective adiabat between models is due to
the differing contribution of radiation pressure.
\label{fig:convectfig}}
\end{figure}

We compute convective models for a range of fluxes and base
temperatures for two of the settling solutions of Figure
\ref{fig:settle}. The first is for the mixed H/He ignition model with
$\dot m=0.1\,\dot m\edd$ and $Z_{\rm CNO}=0.01$. Figure
\ref{fig:convectfig}(a) shows temperature profiles of convective
models with flux $F/F\edd=0.31, 0.74, 0.89$ and $1.15$, the fraction
of the accumulated matter that becomes part of the convective zone,
$(y_b-y_c)/y_b=0.5, 0.8, 0.99$ and $0.95$, and base temperature
$T_b/10^9\,{\rm K}=1.2, 1.5, 1.7$ and $1.7$ respectively. Here we
measure flux in units of the Eddington flux (for the accreted solar
composition) $F\edd=GM\dot m\edd/R$, giving $F\edd\equiv 1.67\times
10^{25}\,{\rm erg\,cm^{-2}\,s^{-1}}$.  Figure \ref{fig:convectfig}(b)
shows convective models corresponding to the pure He ignition settling
solution with $\dot m=0.015\,\dot m\edd$ and $Z_{\rm CNO}=0.01$. We
make convective models with flux $F/F\edd=0.35, 0.59, 1.06$ and
$1.21$, the fraction of the accumulated matter that becomes part of
the convective zone, $(y_b-y_c)/y_b=0.5, 0.8, 0.99$ and $0.95$, and
base temperature $T_b/10^9\,{\rm K}=1.3, 1.6, 2.0$ and $1.95$
respectively. Properties of all these models are given in Table
\ref{tab:conv}. In each case, we take the composition profile of the
atmosphere to be that immediately prior to ignition, but assume full
mixing of matter in the convective zone (we average over the
convection zone according to $\bar{X_i}\equiv \int dy X_i(y)/(y_b-y_c)$
for each species $i$).

We give the thickness of the convection zone $\Delta z_c$ in Table
\ref{tab:conv}, as well as the height above the base which contains
90\% of the mass, $\Delta z(90\%)$.  A simple analytic estimate of the
thickness, $\Delta z_c$, is obtained as follows. First, the adiabatic
index of the gas is
\begin{equation}\label{eq:delad}
n\equiv\left({d\ln T\over d\ln P}\right)_{s}= {8-6\beta\over
32-24\beta-3\beta^2},
\end{equation}
(Clayton 1983), where the ratio of gas pressure to total pressure is
\begin{equation}
\beta=1-0.044\left({T\over 10^9\,{\rm K}}\right)^4 \left({y_b\over
3\times 10^8\,{\rm g\,cm^{-2}}}\right)^{-1} \left({g_{14}\over
1.9}\right)^{-1}.
\end{equation}
For $\beta=1$ ($P=P_g$), we obtain the ideal gas value $n=2/5$;
for $\beta=0$ ($P=P_r$), $n=1/4$. We take $P\propto
T^n$, and presume $\beta$ is a constant throughout the
convective zone, so that $n$ is a constant, given by equation
(\ref{eq:delad}). Integrating $dz=-dP/\rho g$, we find the thickness
of the convection zone is
\begin{equation}\label{eq:zp}
\Delta z_c\approx {H_b\over n}
\left[1-\left({P_t\over P_b}\right)^n\right],
\end{equation}
when it extends to a pressure $P_t$. We denote the scale height at
the base as $H_b\equiv P_b/\rho_bg=k_BT_b/\mu m_pg\beta$ and assume
$\mu$ is constant, giving
\begin{equation}\label{eq:zc}
\Delta z_c\approx 11\,{\rm m}\ {T_{9,b}\over \mu\beta}
\left({0.4\over n}\right) \left({1.9\over
g_{14}}\right)\nonumber\\\left[1-\left({P_t\over
P_b}\right)^n\right].
\end{equation}
Using the value of $\beta$ at the base (denoted $\beta_b$ in Table
\ref{tab:conv}) to evaluate equation (\ref{eq:zc}), we find this
analytic estimate agrees to better than 5\% with the numerical
integrations. Equation (\ref{eq:zc}) shows that radiation pressure
strongly affects the radial extent of the convective zone, $\Delta
z_c\propto 1/\beta$.

\begin{figure}
\plottwo{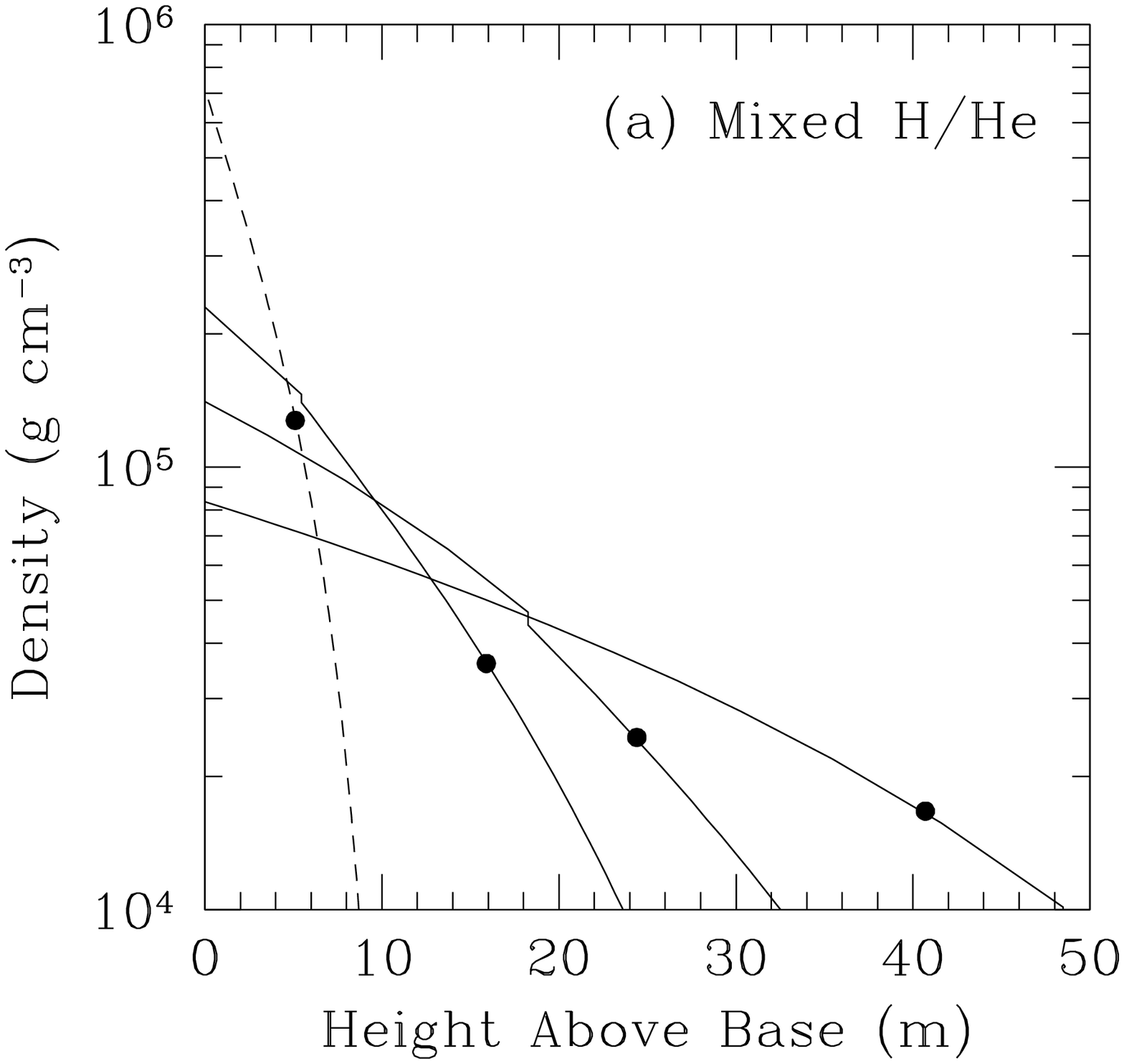}{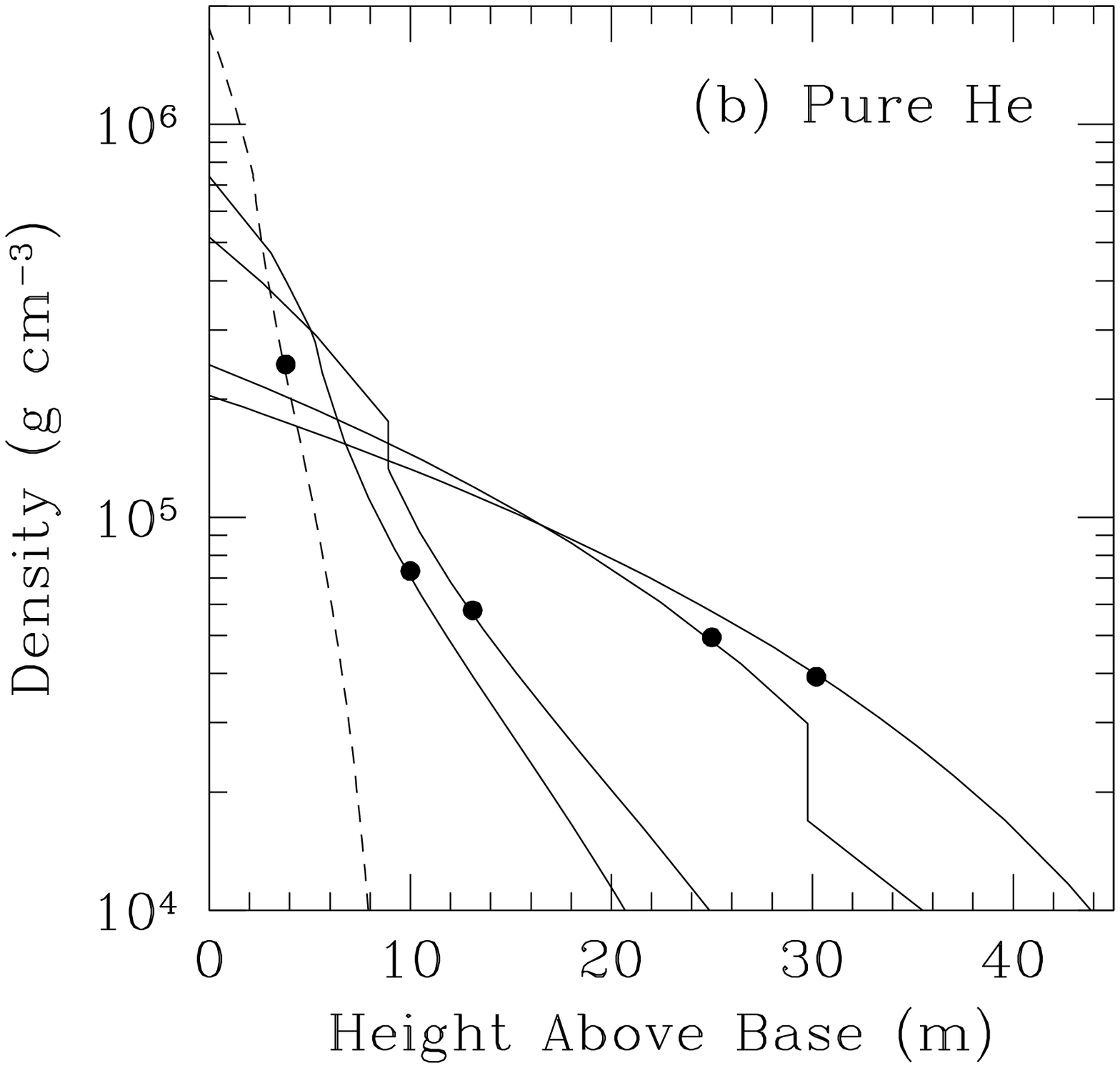}
\caption{Density as a function of height (solid lines) for the
convective models of Table 3 and Figure \ref{fig:convectfig} for (a)
mixed H/He ignition and (b) pure He ignition. The density profile just
before ignition is shown as a dashed line in each case. The black dots
mark the height which encloses 90\% of the mass. In (a) the density
profiles of the two models with $F/F\edd=0.89$ and $1.15$ overlap.
\label{fig:rhoplot}}
\end{figure}

We show the density as a function of height above the base for the
mixed H/He and pure He ignition models in Figure \ref{fig:rhoplot}(a)
and (b) respectively.  The jump in density at the interface between
the convective and radiative zones is due to the jump in composition
because the convective zone is fully mixed. The black dots mark the
height which encloses 90\% of the mass as found numerically. This
ranges from $\approx 15$--$50\ {\rm m}$ for the mixed H/He models, and
$\approx 10$--$30\ {\rm m}$ for the pure He models (Table
\ref{tab:conv}), compared to $\approx 5\ {\rm m}$ before the burst
(Table \ref{tab:ign}). The difference between the mixed H/He and pure
He models arises because $\Delta z_c$ depends strongly on the mean
molecular weight (eq. [\ref{eq:zc}] gives $\Delta z_c\propto
1/\mu$). For a mixture of H and He, $\mu=4/(8X+3Y)$, giving $\mu=4/3$
for pure He, but $\mu=0.6$ for solar composition, almost a factor of
two different. The mean molecular weight for each model is given in
Table \ref{tab:conv}.

\subsubsection{Fully-Radiative Models}\label{sec:radmodels}

We consider two different kinds of radiative models: a radiative
atmosphere carrying constant flux, and a radiative atmosphere with a
constant energy production rate $\epsilon$. The first case is relevant
when the energy production is concentrated near the base, for example
in the initial He burning stages of a burst which does not
convect. The second case is relevant when the burning region is more
extended. For example, in bursts with mixed H/He ignition, the helium
burns rapidly (by purely strong reactions) during the initial stages
of the burst, whereas the hydrogen burns more slowly (by the
rp-process involving weak reactions; Wallace \& Woosley 1981), giving
rise to the long tails seen in some X-ray bursts (for a recent
example, see Figure 2 of Kong et al. 2000). Because the timescale for
H burning by the rp process ($\sim 10$--$100\ {\rm s}$) is long
compared to the time for radiative heat transport ($\sim 1$--$10\ {\rm
s}$), the atmosphere is radiative during the rp process tail of the
burst (Hanawa \& Sugimoto 1982).

\begin{figure}
\plottwo{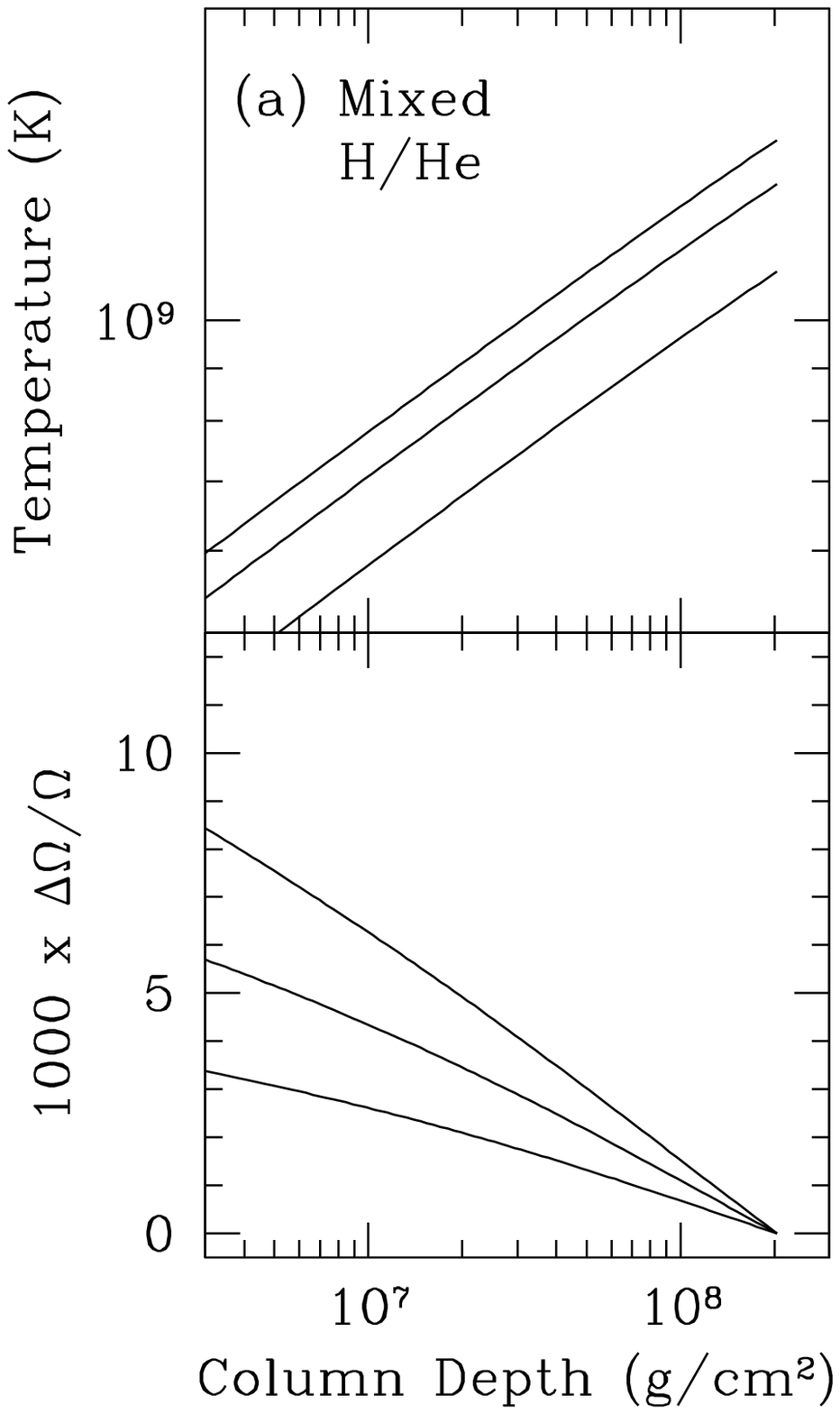}{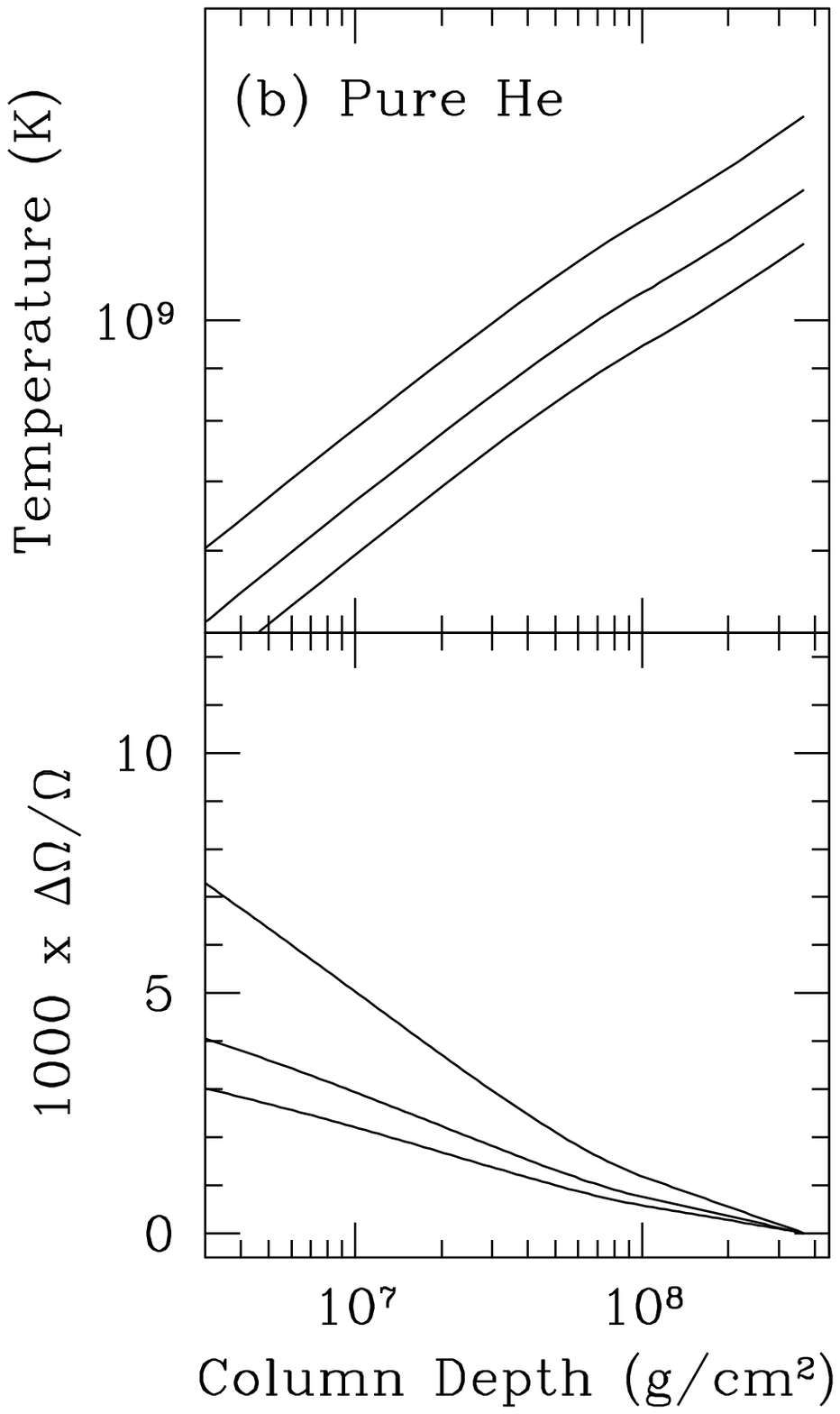}
\caption{Temperature profiles (upper panel) and spin down (lower
panel) for the fully-radiative models of Table \ref{tab:noconv}. (a)
We show constant flux models for mixed H/He ignition with $\dot
m=0.1\,\dot m\edd$, $Z_{\rm CNO}=0.01$, and $y_b=2.04\times 10^8\,{\rm
g\,cm^{-2}}$. These models have $F/F\edd=0.31, 0.74$ and $1.15$ and
$T_b/10^9\ {\rm K}=1.12, 1.35$ and $1.49$. (b) We show constant flux
models for pure He ignition with $\dot m=0.015\,\dot m\edd$, $Z_{\rm
CNO}=0.01$, and $y_b=2.53\times 10^8\,{\rm g\,cm^{-2}}$. These models
have $F/F\edd=0.35, 0.59$ and $1.21$ and $T_b/10^9\ {\rm K}=1.19,
1.34$ and $1.57$.
\label{fig:noconvectfig}}
\end{figure}

\begin{figure}
\plottwo{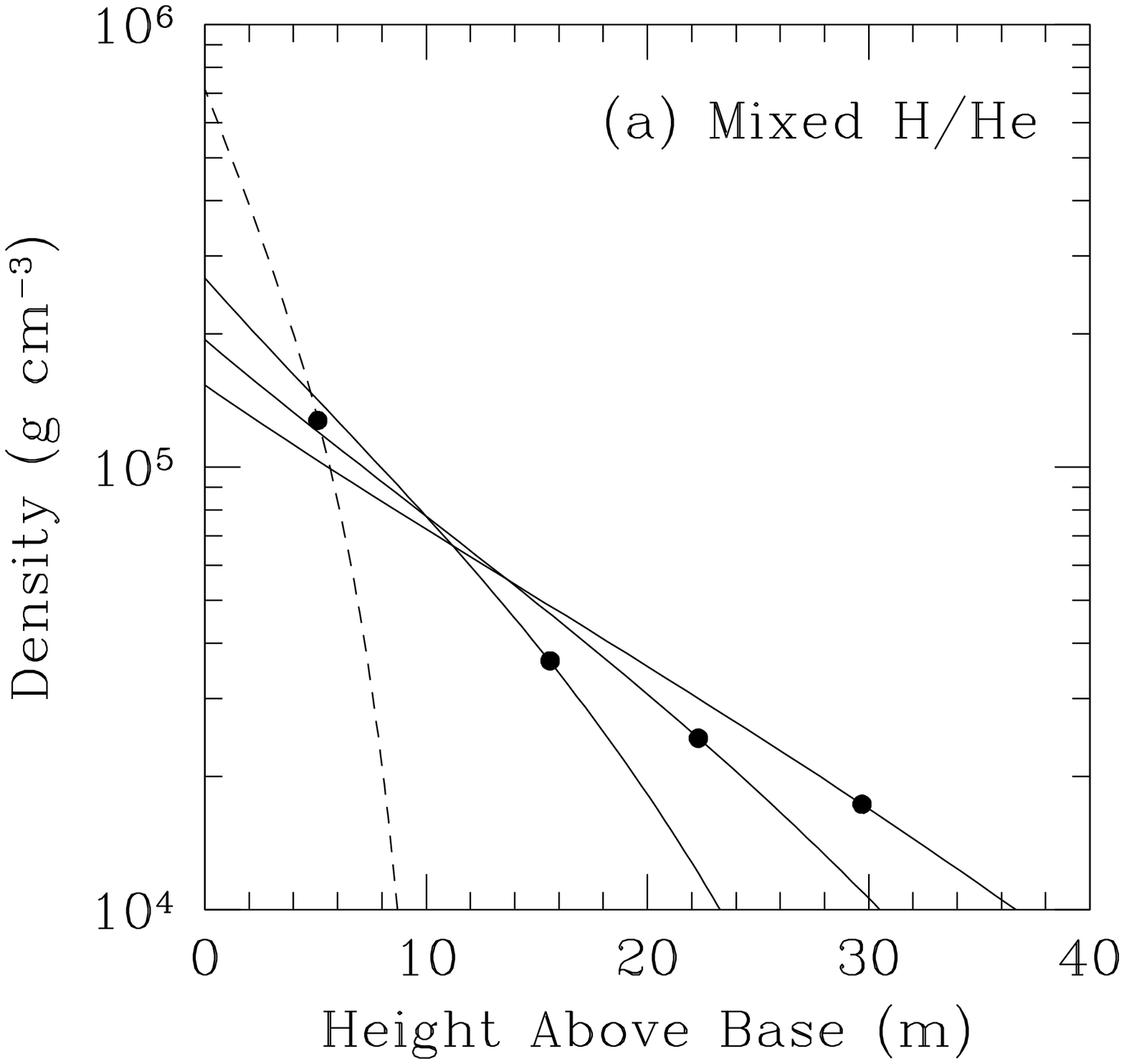}{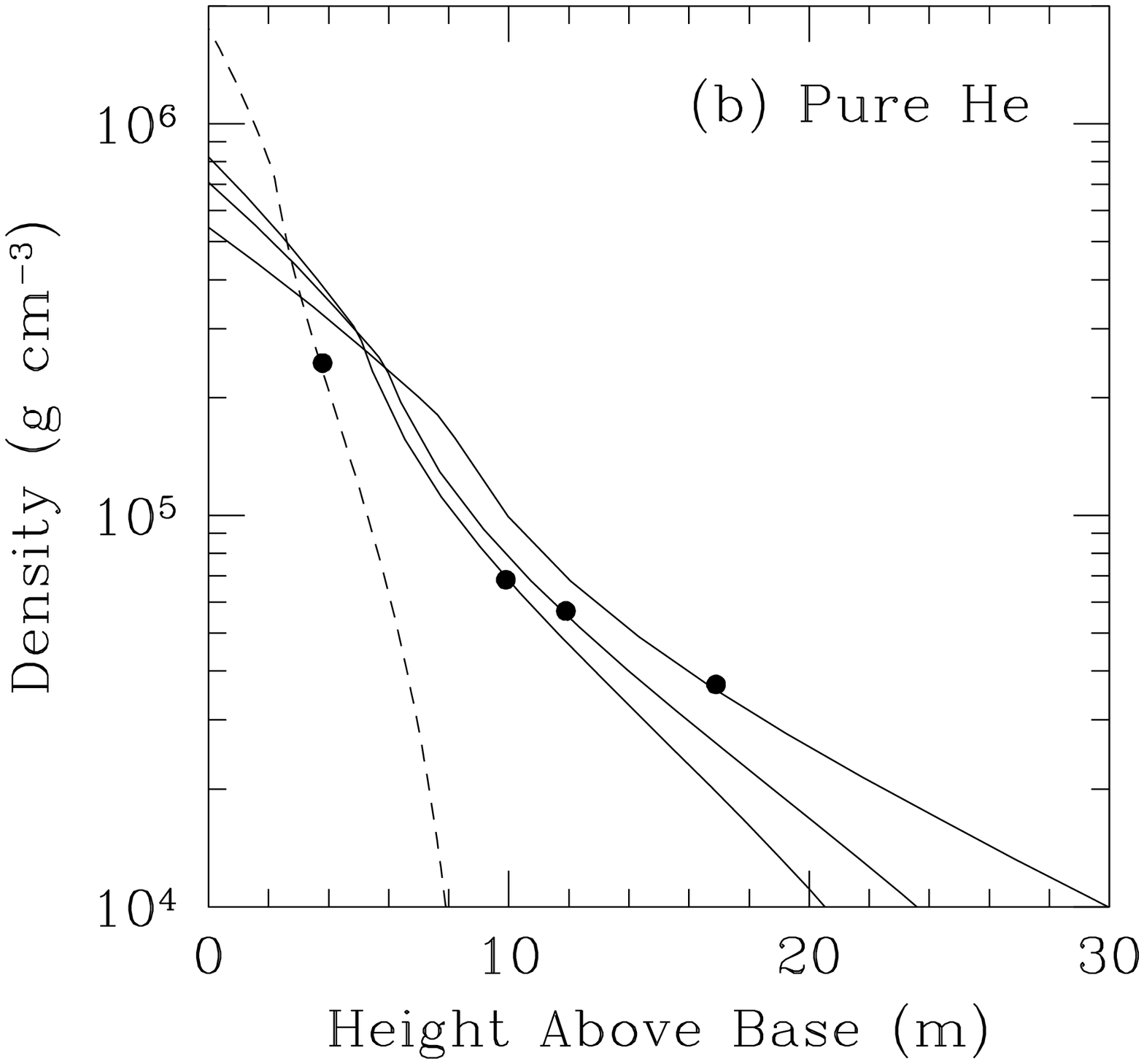}
\caption{Density as a function of height (solid lines) for the
fully-radiative models of Table 4 and Figure \ref{fig:noconvectfig}
for (a) mixed H/He ignition and (b) pure He ignition. The density
profile just before ignition is shown as a dashed line in each
case. The black dots mark the height which encloses 90\% of the mass.
\label{fig:norhoplot}}
\end{figure}

We give details of our fully-radiative models in Table
\ref{tab:noconv}. For the constant $\epsilon$ models, we choose
$\epsilon$ to give the required flux at the top, $F=\epsilon
y_b+F_b$. We assume the composition of the atmosphere is the same as
immediately prior to ignition. Figure \ref{fig:noconvectfig} shows the
temperature profiles. We compute models with $F/F\edd=0.31, 0.74$ and
$1.15$ (mixed ignition) and $F/F\edd=0.35, 0.59$ and $1.21$ (pure He
ignition). Figure \ref{fig:norhoplot} shows the density profiles for
these models. For the mixed ignition models, $\Delta z($90\%$)\approx
15$--$33\ {\rm m}$; for the pure He models, $\Delta z($90\%$)\approx
9$--$17\ {\rm m}$. Again, the pure He models are less extended because
of the larger $\mu$.

\section{Spin Evolution of the Burning Layers}\label{sec:expansion}

We have shown that the atmosphere expands outwards by $\approx 10$--$50\
{\rm m}$ ($\approx 5$--$25\ {\rm m}$) during a mixed H/He (pure He)
burst. In this section, we first calculate how the spin of the burning
layers evolves during the burst, assuming they conserve angular momentum
as they hydrostatically expand and contract, and that they remain
decoupled from the bulk of the neutron star (\S 3.1--3.3). In \S
\ref{sec:heattrans}, we point out (in agreement with Miller 2000) that
for the observer to see a coherent oscillation requires either a
mechanism to enforce rigid rotation of the burning layers (such as
convection), or that the burning layers be geometrically thin. In
addition, we consider the heat propagation through the radiative layers
to the surface, and discuss how this can ``wash-out'' any coherent pulse
emanating from deeper layers. We conclude in \S \ref{sec:radiusexp} by
discussing what happens during radius expansion bursts.

\subsection{Expansion and Spin Down}

We assume that the action of hydrodynamic instabilites (Fujimoto 1988,
1993) or a weak poloidal magnetic field will force the accumulating
pile of fuel to be rigidly rotating with the spin frequency of the
star, $\Omega_\star$. The time for angular momentum to diffuse across
a scale height $H$ due to molecular viscosity $\nu$ alone is
$t_{visc}\approx H^2/\nu$. In \S 5, we show that this time is
$t_{visc}\approx 38\,{\rm h}\,(H/5\ {\rm m})^2$ for $T_8\approx 2$,
comparable to the time between bursts. Thus, even microphysical
mechanisms might bring the accumulated fuel into corotation.

We further assume that the burning is spherically symmetric, ignoring
the complicated, and important, question of how the burning spreads
over the stellar surface, and the fact that some asymmetry is needed
to give an observable oscillation. Thus we assume the angular velocity
is constant on spherical shells, and depends on radius only. This
remains true as the atmosphere expands, since a rigidly rotating
spherical shell stays rigidly rotating as it expands outwards (because
the fractional change in the distance from the rotation axis is the
same for all latitudes). Conservation of angular momentum demands that
$r^2\Omega(r)$ remains constant as a fluid element expands outwards,
giving the spin frequency as a function of column depth during the
burst
\begin{equation}\label{eq:omegarad}
\Omega(y)=\Omega_\star\left({R+z_1(y)\over R+z_2(y)}\right)^2,
\end{equation}
where $z_1(y)$ ($z_2(y)$) is the height of column $y$ above the base
before (during) the burst. We take the base at $y_b$ to be at fixed
radius $R$, even though the old ashes below heat up and expand a
little.  We ignore the general relativistic corrections to the
frequency shift, which Strohmayer (1999a) estimates are a
$\approx$10\%--20\% effect.

Equation (\ref{eq:omegarad}) shows that the spin frequency depends on
depth in the radiative layers. The picture is different in a
convective zone. We expect the convective motions will rapidly bring
the convection zone into rigid rotation, since the convective turnover
time is very short ($\sim 10^{-3}\,{\rm s}$). Thus, if there is no
angular momentum transport between the convective zone and neighboring
layers, the spin frequency of the convective zone is
\begin{equation}
\Omega_c=\Omega_\star\left({I_1\over I_c}\right), 
\end{equation}
where $I_1$ is the total moment of inertia of the material between
$y_c$ and $y_b$ before the ignition, and $I_c$ is the moment of
inertia of the convection zone. We calculate the moment of inertia as
follows. A spherical shell at radius $r$, mass $dm=4\pi r^2\rho\,dr$
has a moment of inertia $dI=(8\pi/3)\rho(r)r^4\,dr$, where the
$8\pi/3$ comes from integrating over angles. Since $\rho dr\equiv
-dy$, this is $dI=-(8\pi/3) r^4 dy$. Thus the moment of inertia is
\begin{equation}\label{eq:I} I(y)= {8\pi R^4\over 3}\int_{y_b}^{y} dy
\left(1+{z(y)\over R}\right)^4,
\end{equation}
where $z(y)$ is the height above the base, given by integrating
$dz/dy=-1/\rho$. We take $P=gy$ during this integration, neglecting
the small variation of $g$ with depth, which has an ${\cal O}(H/R)^2$
effect on the moment of inertia.

We derive a simple analytic estimate of the moment of inertia of the
convective zone, by assuming $\beta=P_g/P$ is constant, so that $z(P)$
is given by an equation similar to equation (\ref{eq:zp}). Inserting
this into equation (\ref{eq:I}), writing $(1+z/R)^4\approx 1+4z/R$ and
integrating, we find
\begin{equation}\label{eq:Ic}
I_c={2\over 3}\Delta M_c R^2\left[1+
{4H_b\over nR}\left(1-{1\over
n+1}{1-x^{n+1}\over 1-x}\right)\right],
\end{equation}
where $x\equiv y_c/y_b$, and $\Delta M_c\equiv 4\pi R^2(y_b-y_c)$ is the
mass of the convection zone. The moment of inertia has two pieces. The
zeroth order piece, $2\Delta M_c R^2/3$ is the moment of inertia of mass
$\Delta M_c$ concentrated at radius $R$. The second term is the ${\cal
O}(H/R)$ correction to this because the envelope is extended. Equation
(\ref{eq:Ic}) reproduces the numerical results to a few percent.

The lower panels of Figure \ref{fig:convectfig} show
$\Delta\Omega/\Omega(y)\equiv (\Omega_\star-\Omega(y))/\Omega_\star$
as a function of column depth for the different convective models in
Table \ref{tab:conv}. We define $\Delta\Omega/\Omega$ so that a
positive value means that the layers have spun down. Figure
\ref{fig:noconvectfig} shows $\Delta\Omega/\Omega(y)$ for the fully
radiative models of Table \ref{tab:noconv}. The values of
$\Delta\Omega/\Omega$ that we find are similar to the observed
frequency shifts during bursts (Table 1). As we described above, in
radiative zones the spin down depends on height, whereas convective
zones are rigidly rotating. In the convective models, this gives rise
to jumps in $\Delta\Omega/\Omega$ at the boundaries between the
convection zone and both the underlying ashes and the overlying
radiative layer. For both convective and fully-radiative models, the
$\Delta\Omega/\Omega(y)$ is smaller by roughly a factor of two for
pure He ignition as opposed to mixed H/He ignition. This is because
the pure He models expand outwards less, as we discussed in \S 2.2.

\subsection{Shrinking of the Convection Zone}

As we described in \S \ref{sec:convmodels}, the atmosphere does not
remain convective for the whole duration of the burst. Eventually, the
convection zone shrinks back. What is the profile of
$\Delta\Omega/\Omega$ left by the retreating convection zone? Imagine
that the convection zone retreats by an amount $d y_c$ (so the top of
the convection zone moves from column $y_c$ to $y_c+d y_c$). The
material which decouples from the convection zone takes away an amount
of angular momentum given by $d J_c=-8\pi (R+\Delta z_c)^4\Omega_cd
y_c/3$, where $\Omega_c$ is the spin frequency of the convection zone,
and $\Delta z_c$ is the height of the top of the convection zone above
the base. The loss of angular momentum from the convection zone must
be compensated by a change in its spin or moment of inertia $d
J_c=d(I_c\Omega_c)=I_cd\Omega_c+\Omega_cdI_c$. Dividing by $dy_c$, and
changing variables to $x\equiv y_c/y_b$, we obtain the differential
equation\footnote{Since $I_c$ depends on both $H_b$ and $x$, one can
write $dI_c/dx=\partial I_c/\partial x+(dH_b/dx)\partial I_c/\partial
H_b$. In fact, the first term in equation (\ref{eq:shrink}) exactly
cancels the $\partial I_c/\partial x$ piece of the second term, giving
$d\Omega_c/dx=(\Omega_c/I_c)(dH_b/dx)\partial I_c/\partial H_b$. Thus
if the convection zone shrank at fixed $H_b$, it would not change its
spin frequency because the angular momentum taken away by the
decoupled fluid exactly cancels the change in the moment of inertia
due to the change in $x$.}
\begin{equation}\label{eq:shrink}
{d\Omega_c\over dx}=-{\Omega_c\over I_c}\left[
{8\pi R^4y_b\over 3}\left(1+{\Delta z_c\over R}\right)^4-
{dI_c\over dx}\right].
\end{equation}
We integrate this equation from the starting values of
$\Omega_c$ and $x$, towards $x\rightarrow 1$ as the convection zone
vanishes. There is one difficulty, however, which is that $\Delta z_c$
and $I_c$ are functions of not only $x$ but also the scale height at
the base $H_b$. If we assume the composition $\mu$ does not change,
then $H_b$ only depends on $T_b$. We model $T_b(x)$ as follows,
\begin{equation}\label{eq:linearT}
T_b(x)=T_0 + \Delta T\left({1-x\over 1-x_i}\right),
\end{equation}
where $T_0$ is the final temperature after the convection vanishes
($x\rightarrow 1$), $T_0+\Delta T$ is the initial temperature at the
base of the convection zone, and $x_i$ is the initial value of $x$. In
reality, $x$ is determined by the instantaneous values of $F$ and
$T_b$, so if we knew $F(T_b)$ we could map this onto
$T_b(x)$. Instead, we specify $T_b(x)$ then as a check that our choice
is reasonable, we can compute $F(T_b)$.

\begin{figure}
\epsscale{0.5}\plotone{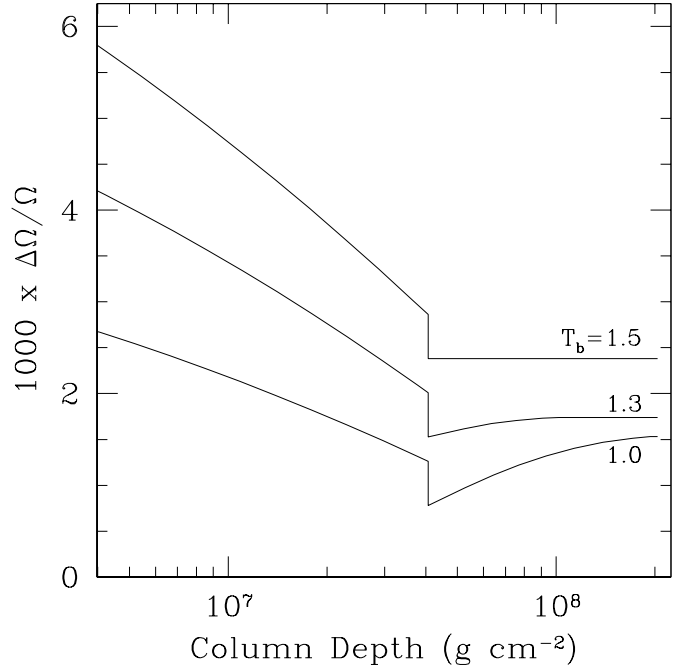}\epsscale{1.0}
\caption{The $\Delta\Omega/\Omega$ profiles at different times as the
convection zone shrinks, for mixed H/He ignition with $\dot m=0.1\dot
m\edd$ and $Z_{\rm CNO}=0.01$. We take the initial extent of the
convection zone to be $(y_b-y_c)/y_b=0.8$ and the base temperature
$T_b/10^9\,{\rm K}=1.5$. As the convection zone retreats, we assume
the base temperature decreases linearly (eq. [\ref{eq:linearT}]) to
$T_b/10^9\ {\rm K}=1.0$. We show $\Delta\Omega/\Omega$ as a function
of depth for the initial model (80\% of mass convective), an
intermediate case (50\%) and when the convective zone has almost
vanished (5\%). These models have $T_b/10^9\,{\rm K}=1.5, 1.31,$ and
$1.03$ and $F/F\edd=0.74, 0.49,$ and $0.21$ respectively.
\label{fig:shrink}}
\end{figure}

Figure \ref{fig:shrink} shows the results for the model with $\dot
m=0.1\dot m\edd$, $Z_{\rm CNO}=0.01$ (mixed H/He ignition), with initial
extent of the convective zone $(y_b-y_c)/y_b=0.8$, and base temperature
$T_b/10^9\,{\rm K}=1.5$. As the convection zone retreats, we assume the
base temperature falls to $T_b/10^9\,{\rm K}=1.0$. We show
$\Delta\Omega/\Omega$ profiles for the initial model (80\% of mass
convective), an intermediate case (50\%) and when the convective zone
has almost vanished (5\%). These models have $T_b/10^9\,{\rm K}=1.5,
1.31,$ and $1.03$ and $F/F\edd=0.74, 0.49,$ and $0.21$ respectively. The
jumps in $\Delta\Omega/\Omega$ at the top and base of the convection
zone persist as it shrinks. In addition, the matter that becomes
radiative spins up as the physical thickness of the underlying
convection zone decreases. This gives rise to an inversion in
$\Delta\Omega/\Omega$, as shown in Figure \ref{fig:shrink}. One might
worry that such an inverted profile would be unstable to axisymmetric
perturbations. However, all the profiles in Figure \ref{fig:shrink} have
increasing specific angular momentum with radius. In addition, the \BV
frequency is so large, $N^2\gg\Omega^2$ (\S \ref{sec:KH}), that we
expect the Solberg-Hoiland criterion for stability (see e.g. Endal \&
Sofia 1978) will always be satisfied.

\subsection{Spin Up in the Cooling Tail}

\begin{figure}
\epsscale{0.5}\plotone{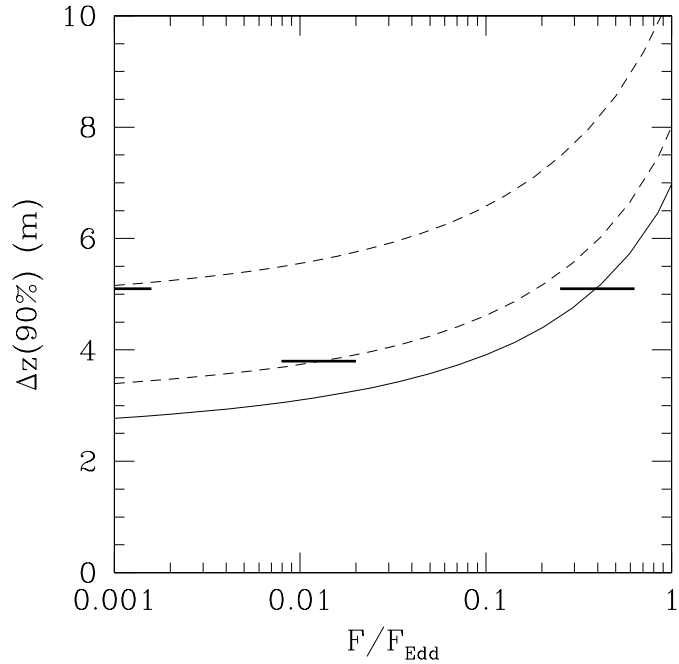}\epsscale{1.0}
\caption{Models of the cooling atmosphere in the tail of the burst. We
show the physical thickness which encompasses 90\% of the mass $\Delta
z($90\%$)$, as a function of the flux $F/F\edd$ for constant flux
atmospheres. The solid curve is for the $\dot m=0.1\ \dot m\edd$,
$Z_{\rm CNO}=0.01$ mixed H/He ignition model, where we take the
composition to be pure $^{73}$Kr. The dashed lines are for pure helium
ignitions, $\dot m=0.01\ \dot m\edd$, $Z_{\rm CNO}=0.01$ (upper curve)
and $\dot m=0.015\ \dot m\edd$, $Z_{\rm CNO}=0.01$ (lower curve). In
this case, we take the composition to be $^{56}$Ni. In each case, we
show $\Delta z(90$\%$)$ just before ignition by a horizontal solid
line.
\label{fig:cool}}
\end{figure}

As the atmosphere cools in the tail of a burst, its thickness
decreases, and it spins back up. We compute simple models of the
cooling atmosphere by assuming it carries a constant flux and has a
fixed uniform composition. In Figure \ref{fig:cool} we show the
thickness $\Delta z($90\%$)$ as a function of the flux $F/F\edd$. The
solid curve is for the $\dot m=0.1\ \dot m\edd$, $Z_{\rm CNO}=0.01$
mixed H/He ignition model, for which we assume a composition of
$^{73}$Kr, since the rp process makes elements beyond the iron group
(Schatz et al. 1998 and references therein; Koike et al. 1999). The
dashed lines are for pure He ignitions, with $Z_{\rm CNO}=0.01$, $\dot
m=0.01\ \dot m\edd$ (upper curve) and $\dot m=0.015\ \dot m\edd$
(lower curve), and for a composition $^{56}$Ni. In each case, we show
$\Delta z(90$\%$)$ just before ignition by a horizontal solid line.

Figure \ref{fig:cool} shows that in the limit $F/F\edd\rightarrow 0$,
the thickness of the ashes is about half the pre-burst thickness for
mixed H/He ignition, but similar to the pre-burst thickness for pure
He ignition. This is because electrons (which provide most of the
pressure support) are consumed in hydrogen burning, but not during
helium burning. In the tail of a burst, when, for example
$F/F\edd\approx 0.1$, the thickness of the atmosphere is $\approx 1\
{\rm m}$ different from the pre-burst thickness in both cases.  Thus
if the burning layers remain decoupled as they cool, we expect the
spin frequency in the tail to be different from the stellar spin by a
part in $\approx 10\ {\rm km}/1\ {\rm m}\approx 10^4$. We discuss the
implications of this result in \S \ref{sec:recouple}.

\subsection{Heat Transport and Coherence of the Oscillation}
\label{sec:heattrans}

We now turn to the vertical transport of heat, and its effect on the
coherence and amplitude of the burst oscillations. Figures
\ref{fig:convectfig} and \ref{fig:noconvectfig} show that in a radiative
zone the magnitude of the spin down depends on depth. However, a single
coherent frequency is observed during bursts. This implies that either
convection or some other mechanism (see \S \ref{sec:coupling}) is
enforcing rigid rotation in the burning region, or, as pointed out by
Miller (2000), the burning layers must be geometrically thin. Since the
fractional change in spin across a height $\Delta z$ in a radiative zone
is $\approx 2\Delta z/R$, an observed coherence $Q$ demands a burning
layer thickness $\lesssim R/2Q$. Strohmayer \& Markwardt (1999) found
$Q\approx 4000$ for bursts from 4U 1728-34 and 4U 1702-43 when the
observed frequency evolution was accounted for, while Muno et al. (2000)
found $Q\approx 5000$ for burst oscillations from KS 1731-26. These
results imply a burning region thickness $\lesssim 100\ {\rm cm}$. This
is perhaps more likely for pure He bursts, because of the temperature
sensitivity of He burning reactions.

Also important is heat transport across the differentially-rotating
atmosphere. The time for radial heat transport from the burning layers
to the photosphere can be estimated from the entropy equation,
$c_P\partial T/\partial t=\partial F/\partial y$, together with equation
(\ref{eq:heat}) for the flux $F$, which gives a timescale $t_{\rm
therm}=3\kappa c_Py^2/4acT^3$. In our numerical calculations, we
calculate the heat capacity at constant pressure $c_P$ exactly. We make
an analytic estimate for a mixture of ideal gas and radiation using
$c_P=(5k_B/2\mu m_p)f(\beta)$ where
$f(\beta)=(32-24\beta-3\beta^2)/5\beta^2$ (Clayton 1983), giving
\begin{equation}\label{eq:ttherm}
t_{\rm therm}=0.6\,{\rm s}
{f(\beta)\over\mu}
\left({\kappa\over 0.08\,{\rm cm^2\,g^{-1}}}\right)
{y_8^2\over T_9^3}.
\end{equation}
The factor $f(\beta)$ is unity for $\beta=1$ ($P=P_g$) and grows
increasingly larger as $\beta$ approaches zero ($P=P_r$), for example
$f(0.85)=2.6$ whereas $f(0.4)=27$ (this is because the internal energy
of a photon gas at constant pressure is independent of
temperature). Thus at a fixed pressure, the thermal time first
decreases with increasing temperature (since $t_{\rm therm}\propto
1/T^3$) but starts increasing again once radiation pressure becomes
important.

Equation (\ref{eq:ttherm}) shows that the thermal time ($\sim 1\ {\rm
s}$) is much longer than the time for hydrostatic readjustment ($\sim
10^{-6}\ {\rm s}$), explaining why spin down is not seen in the
beginning of a burst. By the time the heat released by nuclear burning
reaches the observer, the layers have expanded and spun down.

The burning layers revolve around the underlying star in a time
$\Delta\nu^{-1}\equiv 2\pi/\Delta\Omega\approx 1\ {\rm s}$, comparable to
the thermal time. The ratio of the shearing time to the thermal time
is important. If the heat diffuses quickly, $t_{\rm therm}\Delta\nu\ll
1$, the atmosphere simply transmits the burning pattern from below,
much like passing a flashlight behind a piece of paper. If, on the
other hand, heat diffuses slowly, the oscillations will be smeared
out. In the limit $t_{\rm therm}\Delta\nu\gg 1$, the atmosphere is
heated uniformly by the quickly revolving burning layers.

Figure \ref{fig:therm} shows $t_{\rm therm}$ (solid line) and
$\Delta\nu^{-1}$ (hatched region for $\nu_s=300$--$600\ {\rm Hz}$) as
a function of depth for the convective models of Table
\ref{tab:conv}. For mixed H/He and pure He ignitions, the wrap around
time $\Delta\nu^{-1}$ increases for decreasing base temperature and
physical thickness. The thermal time, however, decreases with
decreasing temperature at a given depth because radiation pressure
(eq. [\ref{eq:ttherm}]) is less important. Figure \ref{fig:therm}
shows that the mixed H/He ignition bursts have $t_{\rm
therm}\Delta\nu\approx 1$, whereas the pure He ignition bursts have
$t_{\rm therm}\Delta\nu\lesssim 1$. This is because both the spin down
and thermal time are less for the pure helium bursts ($\Delta
z_c\propto 1/\mu$ and $t_{\rm therm}\propto 1/\mu$). As we discuss
further in \S \ref{sec:muno}, Figure \ref{fig:therm} may explain the
lack of burst oscillations seen in bursts with the characteristics of
mixed H/He ignition. For these mixed H/He ignition bursts, $t_{\rm
therm}\Delta\nu\approx 1$ and any oscillations in the flux from deeper
regions can get washed out.

\subsection{Radius Expansion Bursts}\label{sec:radiusexp}

Many of the bursts which have shown oscillations are radius expansion
bursts (Table \ref{tab:obs}), in which super-Eddington luminosities
result in expansion of the photosphere to radii $\sim 20$--$100\ {\rm
km}$ (for a review see Lewin, van Paradijs, \& Taam 1993). Burst
oscillations have not been observed during the peak of radius expansion
bursts, but are sometimes seen during the burst rise, and often during
the tail of the burst, once the photosphere has fallen back to the
stellar radius
$R$ (for an example, see Figure 2 of Strohmayer et al. 1998c which
shows a radius expansion burst from 4U 1636-54).

We did not model radius expansion in \S \ref{sec:expansion}. However,
several of our models have super-Eddington fluxes, and thus are
appropriate for the early stages of these bursts. Because the opacity
deep in the atmosphere is smaller than at the photosphere, the
hydrostatic structure we calculate breaks down only in the very upper
layers near the photosphere. Our radiative models are also appropriate
for the cooling tail of these bursts, when the photosphere has fallen
back to the stellar surface.

Whatever the cause of the asymmetry on the neutron star surface, the
fact that oscillations are not seen during the peak of radius expansion
bursts is as we would expect, given the discussion of \S
\ref{sec:heattrans}. The long thermal time across the extended envelope
during radius expansion (Paczynski \& Anderson 1986), as well as the
fact that the horizontal and vertical lengthscales are similar, will
hide any asymmetry and wash out the oscillation.

\section{Angular Momentum Transport and Recoupling}
\label{sec:coupling}

We have shown that if different layers of the atmosphere conserve their
angular momentum, expansion during the burst results in differential
rotation within the burning layers. Thus as we noted in \S
\ref{sec:heattrans}, the fact that a single coherent frequency is
observed implies that either the burning layers are geometrically thin,
or that they must rotate rigidly. While convection may enforce rigid
rotation in the early stages of a burst, we do not expect the atmosphere
to be convective in the cooling tail (see \S \ref{sec:burstmodels}). In
this section, we investigate mechanisms which might transport angular
momentum between the differentially-rotating burning layers. In
addition, we investigate whether the burning layers can remain decoupled
from the underlying cold ashes for the $\approx 10\ {\rm s}$ duration of
the burst and the implied several phase wraps.

\subsection{Kelvin-Helmholtz Instability}\label{sec:KH}

Shear layers are notoriously unstable to hydrodynamic instabilities,
in particular the Kelvin-Helmholtz instability (Chandrasekhar
1961). The shear may be stabilized by buoyancy, however, if the work
that must be done against gravity to mix up the fluid is greater than
the kinetic energy in the shear. The importance of buoyancy is
measured by the Richardson number, ${\rm Ri}=N^2/(dU/dz)^2$ (for
example, see Fujimoto 1988), where $N$ is the \BV frequency, a measure
of the buoyancy. The Kelvin-Helmholtz instability occurs when ${\rm
Ri}<1/4$ (Chandrasekhar 1961; Fujimoto 1988).

There are two sources of buoyancy in the atmosphere, thermal buoyancy
and buoyancy due to the composition gradient. We write the
\BV frequency $N$ as
\begin{equation}\label{eq:N2}
N^2={g\over H}\left[
{\chi_T\over\chi_\rho}\left(n-{d\ln T\over d\ln y}\right)
+{\chi_{\mu_e}\over\chi_\rho}{d\ln\mu_e\over dz}
+{\chi_{\mu_i}\over\chi_\rho}{d\ln\mu_i\over dz}\right]
\end{equation}
(Bildsten \& Cumming 1998), where $\chi_Q\equiv\partial\ln P/\partial
\ln Q$ with the other independent thermodynamic variables held
constant. The first term of equation (\ref{eq:N2}) is the thermal
buoyancy. For an ideal gas atmosphere carrying a constant heat flux in
which the opacity is Thomson electron scattering opacity, this is
$N_{th}^2=3\mu m_pg^2/20k_BT$ (Bildsten 1998), giving
\begin{equation}
{N_{th}\over 2\pi}\approx{3\times 10^4\ {\rm Hz}\over T_9^{1/2}}
\left({\mu\over 0.6}\right)^{1/2}
\left({g_{14}\over 2}\right).
\end{equation}
We estimate the composition gradient terms as $N^2_{\mu}\approx
g\Delta\ln\mu/H$, giving
\begin{equation}\label{eq:Nmu}
{N_{\mu}\over 2\pi}\approx 7\times 10^4\ {\rm
Hz}\left(\Delta\ln\mu\right)^{1/2} \left({10\ {\rm m}\over
H}\right)^{1/2},
\end{equation}
where we have taken the density scale height $H$ as the lengthscale
over which the mean molecular weight changes by $\Delta\ln\mu\sim{\cal
O}(1)$.

We show the \BV frequency during the burst for two fully-radiative
models in the top panel of Figure \ref{fig:ri1}. We show the mixed
H/He ignition model with $F=0.83\ F\edd$, and the pure He ignition
model with $F=0.59\ F\edd$ (see Table \ref{tab:noconv}). We show the
total buoyancy by a solid line, the thermal buoyancy as a dotted line
and the composition piece as a dashed line. For the pure He ignition
model, there is a peak in the buoyancy at the place where the hydrogen
runs out ($y\approx y_d$), at this depth the composition piece of the
buoyancy dominates the thermal piece. Our models do not include the
composition jump at the base between the burning layers and the
ashes. The point marked by a cross at the base of each model shows an
estimate of the buoyancy at the base, where we estimate the
composition piece of the buoyancy using equation (\ref{eq:Nmu}) and
take the ashes to have $\mu=2.1$ (for a single species, $\mu=A/1+Z$).

The Richardson number ${\rm Ri}=N^2/(dU/dz)^2$ is plotted in the lower
panel of Figure \ref{fig:ri1} for $\nu_s=300$--$600\ {\rm Hz}$. Since
$\Delta\Omega\approx (2z/R)\Omega$, we estimate
$dU/dz=Rd\Delta\Omega/dz\approx 2\Omega$, giving
\begin{equation}\label{eq:Ri}
{\rm Ri}\approx 7000\ \left({\nu_s\over 300\ {\rm Hz}}\right)^{-2}
\left({N\over 5\times 10^4\ {\rm Hz}}\right)^2,
\end{equation}
which agrees well with Figure \ref{fig:ri1}. As suggested by Bildsten
(1998), the strong buoyancy in the atmosphere gives ${\rm Ri}\gg
1$. Thus we do not expect the Kelvin-Helmholtz instability to operate
either within the burning layers or between the burning layer and the
underlying ashes.

\vspace{1 cm}
\subsection{Ekman Pumping}

We now investigate how fast viscosity acts to smooth out differential
rotation. The molecular viscosity in the atmosphere is determined by
ion-ion collisions, giving
\begin{equation}
\nu\approx 88\ {\rm cm^2\,s^{-1}}\,
\left({T_9^{5/2}\over \rho_5}\right)
\left({A^{1/2}\over Z^{4}}\right)
\left({8\over \log\Lambda}\right)
\end{equation}
(Spitzer 1962), where $T_9\equiv T/10^9\,{\rm K}$,
$\rho_5\equiv\rho/10^5\,{\rm g\,cm^{-3}}$ and $\log\Lambda$ is the
Coulomb logarithm. Thus the time for molecular viscosity to transport
angular momentum over a scale height by diffusion is
\begin{equation}
t_\nu\approx 3.2\ {\rm h}\ \left({H\over 10\ {\rm m}}\right)^2
\left({\nu\over 88\ {\rm cm^2\,s^{-1}}}\right)^{-1},
\end{equation}
much longer than a rotation period. In this case, it is possible to
exchange angular momentum on a faster timescale than $t_\nu$ by the
process of Ekman pumping. This mechanism involves a secondary
circulation in which fluid elements are exchanged between the bulk of
the fluid and the thin viscous boundary layer. Ekman pumping is
well-studied in fluid dynamics (Benton \& Clark 1974), and Livio \&
Truran (1987) suggested that it may operate in accreting white
dwarfs. For a non-stratified fluid, the Ekman spin up or spin down
time is $t_E\approx (t_\nu/\Omega)^{1/2}$ or
\begin{equation}\label{eq:Ekman}
t_E\approx 2.5\ {\rm s}\ 
\left({t_\nu\over 3.2\ {\rm h}}\right)^{1/2}
\left({\nu_s\over 300\ {\rm Hz}}\right)^{1/2}
\end{equation}
(Greenspan \& Howard 1963; Benton \& Clark 1974). The thickness of the
viscous boundary layer is $\approx (\nu/\Omega)^{1/2}\approx
0.2\ {\rm cm}\ (\nu/88\ {\rm cm^2\,s^{-1}})^{1/2}(\nu_s/300\ {\rm
Hz})^{-1/2}$.

Spin up in a stratified fluid is different since the buoyancy may
inhibit vertical motion of fluid elements, limiting the extent of the
secondary circulation. Spin up in a cylinder of radius $a$ with $g$
parallel to $\Omega$ was first studied by Holton (1965; see Benton \&
Clark 1974 for a review). In this case, the secondary circulation is
confined to a layer of vertical thickness $\approx a(\Omega/N)$ (Walin
1969). Sakurai, Clark \& Clark (1971) found a similar result for a
spherical stratified fluid, for which Ekman spin up occurs in a thin
layer of radial extent $\approx R(\Omega/N)$.  The neutron star
atmosphere during a burst has $H<R(\Omega/N)\approx 60\ {\rm
m}$. Thus, if the spherical result is applicable to a thin layer on
the surface of a sphere, we would not expect the buoyancy to inhibit
Ekman pumping.  In addition, we have $t_{\rm therm}\approx t_E$, so
non-adiabatic effects may be important, reducing the effect of the
thermal buoyancy (Sakurai et al. 1971). Detailed calculations are
needed to find whether Ekman pumping operates during the burst.

\subsection{Baroclinic Instability}\label{sec:baroclinic}

The baroclinic instability is a hydrodynamic instability in which the
fluid motions are close to horizontal, so it occurs even when the
Richardson number is large. It has been well-studied in geophysics
(Pedlosky 1987) and in astrophysics because of its possible role in
angular momentum transport in stellar interiors (Knobloch \& Spruit
1982; Tassoul \& Tassoul 1982; Spruit \& Knobloch 1984) and accreting
compact objects (Fujimoto 1988, 1993). We start by showing that in the
presence of a vertical shear, the surfaces of constant pressure and
density are inclined with respect to each other. This represents a store
of gravitational energy which the baroclinic instability can tap. We
then present the results of a stability analysis of a plane-parallel
model, for which we find that the strong buoyancy limits the baroclinic
instability to short vertical wavelengths ($\lesssim 200\ {\rm cm}$).

\subsubsection{The Nature of the Baroclinic Instability}

We first study the misalignment of the constant density and pressure
surfaces that arises in the presence of vertical shear on a rotating
star. This result is well-known on the Earth, where differential heating
between the equator and pole gives rise to the ``thermal wind''
(Pedlosky 1987). To investigate recoupling of the ashes and burning
layers, we adopt a simple ``two layer'' model, in which we include only
the buoyancy associated with the interface between the ashes and the
burning layers. Thus, we take the upper (lower) layer to have constant
density $\rho_+$ ($\rho_-$). Both layers are in hydrostatic balance in
the vertical direction. As elsewhere in this paper, we assume the
burning front has spread over the whole surface.

We work in the rotating frame, in which the lower fluid is
stationary. We assume the upper layer is rigidly rotating with angular
velocity $-\Delta\Omega$ (so that $\Delta\Omega$ is positive for a
spun down shell). Since it is moving, the upper fluid feels a Coriolis
force in the transverse direction. In a time
$\approx\Omega^{-1}\approx 10^{-3}\ {\rm s}$, a transverse pressure
gradient will be established which balances the Coriolis force
(geostrophic balance). However, in the lower fluid there is no
transverse pressure gradient (it feels no Coriolis force). Since the
pressure must be continuous across the boundary, it cannot be
horizontal, but must slope.

Figure \ref{fig:slope}(a) illustrates this. We show a small section of
the two layers at latitude $\theta$ ($\theta$ is the angle from the
pole) in the $(r,\theta)$ plane. Because the star is rotating, the
equilibrium isobars are not spherical, but nearly (the stellar radius
is larger at the equator than the pole by $\approx
R^2\Omega^2/g\approx 100\ {\rm m}$ for typical parameters). We adopt
coordinates $z$ and $y$ perpendicular and parallel to surfaces of
constant pressure (dotted line). The vertical component of the
rotation vector is $\Omega\cos\theta$, and the upper fluid is moving
out of the page with velocity $U=R\Delta\Omega\sin\theta$.

Now consider the pressure changes as we move from point A to point B
on the boundary. First we move through the lower fluid along the
dashed line, horizontally a distance $dy$, then vertically
upwards a distance $dz$. The pressure change along this path is
$dP_-=-\rho_{-}g\,dz$. Second, we move through the upper fluid along
the dashed line. This time, there is a pressure change while moving
horizontally also, giving $dP_+=-\rho_{+}g\,dz-\rho_{+}2\Omega
U\cos\theta\,dy$. Demanding $dP_-=dP_+$, gives the slope
\begin{equation}\label{eq:slope}
{dz\over dy}={2\Omega^2R\over
g}{\Delta\Omega\over\Omega}{\rho_{+}\over\Delta\rho}
\cos\theta\sin\theta,
\end{equation}
where $\Delta\rho=\rho_{-}-\rho_{+}>0$. If we take the equilibrium
equipotentials to be spherical ($dy\approx Rd\theta$) and integrate,
we find the change in height of the boundary is
$z_b(\theta)=-\bar{z}_b\cos 2\theta+\,{\rm const.}$, where
$\bar{z}_b=(\Omega^2R^2/2g)(\Delta\Omega/\Omega)(\rho_+/\Delta\rho)$
or
\begin{equation}
\bar{z}_b=35\ {\rm cm}\ \left({\nu_s\over 300\ {\rm
Hz}}\right)^2
\left({\Delta\Omega/\Omega\over 4\times
10^{-3}}\right) \left({\rho_+\over\Delta\rho}\right),
\end{equation}
where we have taken $g_{14}=2$ and $R=10\ {\rm km}$.

The sloping interface represents a store of gravitational potential
energy. Assuming constant density, the displaced mass per unit area is
$\rho z_b(\theta)$, and the gravitational energy per unit area
required to displace the interface is $\rho
gz_b(\theta)^2/2$. Integrating over a spherical surface, we find the
gravitational energy stored in the interface is $\Delta
E_g=(16/45)(4\pi R^2\rho_{+}g\bar{z}_b^2/2)(\Delta\rho/\rho_{+})$, or
\begin{eqnarray}
\Delta E_g=6\times 10^{34}\ {\rm erg}\ 
\left({\Delta\rho\over\rho_+}\right)
\left({\bar{z}_b\over 35\ {\rm cm}}\right)^2\nonumber\\
\left({\rho_+\over 10^5\ {\rm g\ cm^{-3}}}\right)
\left({g_{14}\over 2}\right).
\end{eqnarray}
The amount of displaced mass is $\approx 4\pi R^2\rho\bar{z}_b\approx
4\times 10^{19}\ {\rm g}$, so $E_g$ is only a few keV per nucleon,
much less than the energy produced by nuclear burning. Thus
establishing the sloping interface poses no energetic obstacle. For
comparison, the energy in the shear is $\approx\rho H_b U^2/2$ per
unit area, giving $E_U\approx (8\pi R^2/3)\rho H_b
(R\Delta\Omega)^2/2$ or
\begin{eqnarray}
E_U\approx 3\times 10^{34}\ {\rm erg}\ 
\left({\rho\over 10^5\ {\rm g\ cm^{-3}}}\right)
\left({H_b\over 10\ {\rm m}}\right)\nonumber\\
\left({\nu_s\over 300\ {\rm Hz}}\right)^2
\left({\Delta\Omega/\Omega\over 4\times 10^{-3}}\right)^2,
\end{eqnarray}
when integrated over the surface.

The baroclinic instability acts to release the gravitational potential
energy stored in the misaligned pressure and density
surfaces. Consider the fluid displacements shown in Figure
\ref{fig:slope}(b). If the fluid element is displaced to point A, it
is heavier than its new surroundings and is pushed back by the
buoyancy of the interface. However, if it is moved to point B, it
falls in the gravitational field, releasing energy. Any displacement
within the so-called ``wedge of instability'' (Pedlosky 1987) is
convectively unstable in this way.

\subsubsection{Results of Stability Analysis}

We have carried out a linear stability analysis of a plane parallel
two-layer model. We do not present our detailed calculations, rather we
summarize our results and use simple physical arguments to understand
them. We start by considering motions about a latitude
$\theta=\theta_0$, and adopting a local cartesian coordinate system
$(x,y,z)$ where the transverse coordinate $x$ ($y$) is in the $\phi$
($\theta$) direction. We include the effect of sphericity using the
``beta-plane approximation'' of geophysics (Pedlosky 1987), i.e. we
write
\begin{equation}
2\Omega\sin\theta\approx
2\Omega\sin\theta_0+(\theta-\theta_0)\cos\theta_0,
\end{equation}
where we assume $(\theta-\theta_0)=y/R\ll 1$. We consider a channel
centered on $\theta_0=\pi/4$ and of width $\pi R/4$, stretching from
the pole to the equator. We write the fluid displacement as
($\xi_x$,$\xi_y$,$\xi_z$), and look for solutions $\propto\exp({\sigma
t+ik_xx+ik_yy})$ where $k_x=2\pi/\lambda_x$ and $k_y=2\pi/\lambda_y$
are the transverse wavenumbers, and $\sigma$ is the growth
rate. Pedlosky (1987) performs a similar analysis, although restricted
to the case $\Delta\rho/\rho\ll 1$.

We find that small wavelengths are stable because the fluid
displacements lie outside the wedge of instability. To see this, assume
that the perturbations are in geostrophic balance, in which case the
continuity equation gives $\xi_z/\lambda_z\approx {\rm
Ro}\,\xi_\perp/\lambda_\perp$, where ${\rm Ro}\approx
U/2\Omega\lambda_\perp\approx R\Delta\Omega/2\Omega\lambda_\perp$ is the
Rossby number of the perturbation, and the transverse wavelength
$\lambda_\perp=2\pi/k_\perp$, with $k_\perp^2=k_x^2+k_y^2$. Then the
angle of fluid displacement is $\xi_z/\xi_y\approx\xi_z/\xi_\perp\approx
R\Delta\Omega H/\lambda_\perp^2\Omega$, where the vertical wavelength is
set by the vertical extent of the burning layers $\lambda_z\approx
H$. For instability, the angle of fluid displacement should be less than
the slope of the interface (eq. [\ref{eq:slope}]), requiring
$\lambda_\perp^2\gtrsim (gH_b/\Omega^2)(\Delta\rho/\rho_+)$, or
\begin{eqnarray}\label{eq:lambdabc}
\lambda_\perp>\lambda_{BC}=7.5\times 10^5\ {\rm cm}\ \left({H_b\over
10\ {\rm m}}\right)^{1/2} \left({g_{14}\over
2}\right)^{1/2}\nonumber\\ \left({\Delta\rho\over\rho_+}\right)^{1/2}
\left({\nu_s\over 300\ {\rm Hz}}\right)^{-1},
\end{eqnarray}
where we insert the correct prefactor. Thus the nearly horizontal
unstable displacements require transverse wavelengths of order the
stellar radius or greater. 

However, very large transverse wavelength perturbations can be
stabilized because the vertical component of $\Omega$ changes
significantly across a wavelength. The changing vorticity
provides a restoring force in the $\theta$ direction (Pedlosky
1987). This is the same force which supports Rossby waves (Pedlosky
1987; Brekhovskikh \& Goncharov 1994; Dutton 1995). We find that the
convective instability overcomes the Rossby wave restoring force when
the shear velocity is greater than the Rossby wave speed, $U>U_{\rm
Ro}\approx 2\Omega\lambda_\perp^2/R$ (see also Pedlosky 1987).  Thus
we require $\lambda_\perp\lesssim R(\Delta\Omega/\Omega)^{1/2}$, or
\begin{equation}\label{eq:lambdaro}
\lambda_\perp<\lambda_{\rm Ro}
=3.3\times 10^5\ {\rm cm}\
\left({\Delta\Omega/\Omega\over 4\times 10^{-3}}\right)^{1/2}
\left({R\over 10\,{\rm km}}\right)
\end{equation}
for instability, where we insert the correct prefactor.

Transverse wavelengths which satisfy equation (\ref{eq:lambdabc}) do
not satisfy equation (\ref{eq:lambdaro}). Thus in the context of our
plane-parallel model, we conclude that, when $H=10\ {\rm m}$, large
scale mixing between the burning layers and ashes by the baroclinic
instability does not occur. Instability can occur when $\lambda_{\rm
BC}<\lambda_{\rm Ro}$, or $\Delta\rho/\rho_+\lesssim
(4\Omega^2R/g)(\Delta\Omega/\Omega)(R/H)$. Since
$\Delta\Omega/\Omega\approx 2H/R$, this is
\begin{equation}
{\Delta\rho\over\rho_+}\lesssim 0.15\ \left(\nu_s\over 300\ {\rm
Hz}\right)^2\left({g_{14}\over 2}\right),
\end{equation}
so mixing may occur as the burning layers cool and the density contrast
with the ashes decreases. The growth rate of an unstable mode is
$\propto k_x$, so that writing $k_x\approx m/R$ (so that $e^{ikx}\sim
e^{im\phi}$), we estimate the fastest growing modes will have $m\approx
20\ (\lambda_\perp/3\times 10^5\ {\rm cm}$). Several authors have
studied two layer models on a sphere (Hollingsworth 1975; Hollingsworth,
Simmons \& Hoskins 1976; Simmons \& Hoskins 1976, 1977; Moura \& Stone
1976; Warn 1976), but for parameters of interest for the Earth, namely
$\Delta\rho/\rho\ll 1$ and $\Delta\Omega/\Omega\approx 0.01$. They find
good agreement with growth rates calculated with beta-plane
models. However, further calculations are needed to extend our analysis
to the spherical case, and determine the spherical eigenfunctions and
growth rates.

\subsubsection{Short Wavelength Modes}

So far we have considered only vertical wavelengths $\lambda_z\approx
H$, because of the simple vertical structure of the two layer
model. However, if we allow perturbations with $\lambda_z<H$, we expect
to find instability for small enough vertical wavelengths. This is
because for small vertical wavelengths, the fluid displacement is able
to lie within the wedge of instability (Figure \ref{fig:slope}) while
still having a transverse wavelength which is unaffected by Rossy wave
restoring forces (eq. [\ref{eq:lambdaro}]).

We make a simple estimate of the vertical wavelength at which
instability occurs by repeating the arguments leading to equation
(\ref{eq:lambdabc}), but this time allowing $\lambda_z<H_b$. In this
case, $\lambda_{BC}$ is given by equation (\ref{eq:lambdabc}) with $H_b$
replaced by $\lambda_z$. For instability, we require $\lambda_{\rm
BC}<\lambda_{\rm Ro}$ (eq. [\ref{eq:lambdaro}]) giving
\begin{equation}\label{eq:lambdaz}
\lambda_z\lesssim 200\ {\rm cm}\ \left({\Delta\Omega/\Omega\over 4\times
10^{-3}}\right) \left({\nu_s\over 300\ {\rm Hz}}\right)
\left({\rho_+\over\Delta\rho}\right),
\end{equation}
where we take $g_{14}=2$ and $R=10\ {\rm km}$. We expect modes with
vertical wavelengths small enough to satisfy equation (\ref{eq:lambdaz})
to be unstable.

Fujimoto (1988, 1993) used the short wavelength baroclinic modes to
define a turbulent viscosity $\nu_{\rm turb}\approx \lambda_z^2/t_{\rm
grow}$ where the growth time of the instability is $t_{\rm
grow}\approx\nu_s^{-1}{\rm Ri^{1/2}}$, or
\begin{equation}
t_{\rm grow}\approx 0.25\ {\rm s}\
\left({\nu_s\over 300\ {\rm Hz}}\right)^{-1}
\left({{\rm Ri}\over 7000}\right)^{1/2}
\end{equation}
(Pedlosky 1987; Fujimoto 1988). The time to transport angular momentum
across a scale height is $t_\nu\approx H^2/\nu_{\rm turb}\approx t_{\rm
grow}(H/\lambda_z)^2$, or
\begin{equation}
t_\nu\approx 6\ {\rm s}\ 
\left({t_{\rm grow}\over 0.25\ {\rm s}}\right)
\left({H\over 10\ {\rm m}}\right)^2
\left({\lambda_z\over 200\ {\rm cm}}\right)^{-2}
\end{equation}
Thus turbulent transport of angular momentum driven by the short
wavelength baroclinic modes could be important during a burst. In
particular, these could act to force the burning shell to rigidly
rotate.

\subsection{Magnetic Field Winding}

The majority of neutron stars in LMXBs show no evidence for coherent
pulsations in their persistent emission. This implies that these neutron
stars are weakly magnetic ($B\ll 10^{9}\ {\rm G}$) so the accretion flow
is likely not disrupted before it reaches the neutron star surface. We
now ask what effect would a weak magnetic field have on the shearing
atmosphere during a burst?  The Ohmic diffusion time across a scale
height is $t=4\pi\sigma H^2/c^2$, or
\begin{equation}
t_{\rm diffuse}\approx {10^7\,{\rm s}\over Z}\ \left({H\over 10\ {\rm
m}}\right)^2\left({T\over 10^9\ {\rm K}}\right)^{3/2}
\left({8\over\log\Lambda}\right)
\end{equation}
(Brown \& Bildsten 1998) where we estimate the conductivity as
\begin{equation}
\sigma\approx {2(2k_BT)^{3/2}\over \pi^{3/2}m_e^{1/2}Ze^2\log\Lambda}
\end{equation}
(Spitzer 1962), $Z$ is the mean ionic charge and $\log\Lambda$ is the
Coulomb logarithm. Since $t_{\rm diffuse}\gg 10\ {\rm s}$, the MHD limit
applies during the burst and the shearing atmosphere will bend the
magnetic field lines. An initially poloidal magnetic field would prevent
the shearing if the energy density in the field is greater than the
shear energy, $B^2/8\pi > \rho(R\Delta\Omega)^2/2$, or
\begin{equation}\label{eq:B10}
B> 10^{10}\ {\rm G}\ \rho_5^{1/2}
\left({R\over 10\ {\rm km}}\right)
\left({\Delta\Omega/\Omega\over 4\times 10^{-3}}\right).
\end{equation}
This is a similar limit to those from lack of persistent pulsations
and from spectral modelling (Psaltis \& Lamb 1998). The observed
shearing for ten seconds most likely rules out fields this strong.

What can we say about a weaker, initially poloidal field like that
often discussed for the the progenitors of millisecond pulsars, $B\sim
10^{8}-10^{9} \ {\rm G}$?  Fields this weak can also have a
significant effect when in the ideal MHD limit. In this case, the
horizontal shearing and dragging of the field lines  generates a
strong toroidal field.  Spruit (1999) recently considered the
interaction of magnetic fields and differential rotation in stellar
interiors. He considered an initially poloidal field
$(B_z,B_\theta,0)$, and showed that $B_\phi$ grows with time according
to $B_\phi=NB_z$ where $N$ is the number of windings,
$N=R\sin\theta\int |\grad\Omega| dt$.  We have $\grad\Omega\approx
\Delta\Omega/H$, giving $N\approx (R\Delta\Omega/H)t\approx 2\Omega
t$. Thus we find $B_\phi\approx B_z$ after only a few revolutions of
the star, much less than the wrap around time of the shear ($\approx
1\ {\rm s}$). The toroidal field grows so quickly because the field is
sheared on a vertical scale $H\ll R$. Another way to see this is to
consider a vertical tube of fluid in the shearing region (with height
$H$ and cross-section area $A_i$) that encloses a magnetic flux
$\Phi=B_z A_i$. After one differential wrapping time (roughly one
second), this fluid element will be stretched into a very thin tube of
length $2\pi R$ and so will have shrunk in cross-sectional area to
$A_f=A_i H/2\pi R$. If flux-freezing holds, then after one second,
$B_\phi\approx B_z A_i/A_f\approx B_z 2\pi R/H\approx 10^4 B_z$. 

This simple argument says that an initial field of $\approx 10^6\ {\rm
G}$ can become dynamically important during the burst.  However, further
theoretical investigations are needed to see if the implied wrapping of
the field is possible without instabilities setting in.  If correct,
this result implies that the bursters with frequency drift have magnetic
fields much weaker than the $B\sim 10^{8}$--$10^9\ {\rm G}$ fields
usually assumed.

\subsection{Summary of Coupling Mechanisms}

We have investigated a number of different mechanisms that could enforce
rigid rotation within the burning layers, or act to couple the burning
layers to the underlying ashes. We find that the strong buoyancy of the
atmosphere prohibits the Kelvin-Helmholtz instability and mixing between
the ashes and the burning layers by the baroclinic instability. Short
vertical wavelength baroclinic modes within the burning region may be
unstable and vigorous enough to force the burning layers to rigidly
rotate. The timescale for viscosity to act via an Ekman pumping
mechanism may be a few seconds or less (eq. [\ref{eq:Ekman}]), but the
effect of buoyancy, in particular from the composition gradients, is not
clear. Finally, our initial estimates suggest that winding of a weak
magnetic field by the differential rotation is important during the
burst.

In summary, we have not found a robust hydrodynamical mechanism that
recouples the burning layers to the star during the $\approx 10\ {\rm
s}$ burst duration. It thus seems plausible that the burning layers
remain decoupled throughout the burst. Short wavelength baroclinic
instabilities may operate within the burning layers, although more work
is needed to see whether they cause the burning shell to rigidly rotate.

\section{Discussion}

We now summarize the hydrostatic expansion of the atmosphere during a
Type I X-ray burst, and the implied spin evolution of the decoupled
burning layers. We find the expected spin changes are of the same order
as the observed frequency drifts, thus supporting the picture proposed
by Strohmayer et al. (1997a). We then use our results to address two
observational issues: (i) why oscillations are not seen in all bursts,
and (ii) the observed long term stability of the oscillation frequency.

\subsection{Temporal Evolution of the Burning Shell}\label{sec:temporal}

The atmospheric evolution during the burst is shown schematically in
Figure \ref{fig:scheme} for (a) mixed H/He ignition and (b) pure He
ignition. As revealed by one-dimensional simulations, the rapid energy
release from helium burning reactions makes the initial stages of most
bursts convective. Later in the burst, the convection zone retreats,
leaving a purely radiative atmosphere. Figure \ref{fig:summ} shows the
thickness of the atmosphere which contains 90\% of the mass, $\Delta
z(90$\%$)$, as a function of the flux for constant flux radiative
atmospheres (solid and dotted lines) and for the convective models of
Table \ref{tab:conv} (solid and open squares). For the radiative
models, the upper curve is for a composition profile the same as at
ignition, while the lower curve is for a composition of $^{73}$Kr
($^{56}$Ni) for the mixed H/He (pure He) case, chosen to represent the
products of burning. Initially, the evolution of the thickness of the
layer is along or above the upper curve as it ignites and heats up,
depending on the extent of the convection zone. As nuclear burning
proceeds, the mean molecular weight increases, and the thickness
decreases, eventually moving back along the lower curve as the
atmosphere cools. Just before ignition (upper curve, $F\ll F\edd$),
the thickness of the accumulated layer is $\approx 5\ {\rm m}$. During
the burst, the atmosphere expands hydrostatically by $\Delta z\approx
10$--$40\ {\rm m}$ ($\approx 5$--$20\ {\rm m}$) for mixed H/He (pure
He) ignition. For a given flux, the convective models have a greater
thickness than fully-radiative models because the temperature profile
is steeper in the convection zone, giving a larger base temperature
(compare Tables \ref{tab:conv} and \ref{tab:noconv}). As we showed in
\S \ref{sec:burstmodels}, radiation pressure plays an important role
when $F\approx F\edd$, acting to lower the density and increase the
hydrostatic expansion.

We showed in \S \ref{sec:expansion} that in a radiative atmosphere, the
magnitude of the spin down depends on depth. Thus to observe a single,
coherent frequency requires either the atmosphere to be rigidly
rotating, or the burning region to be vertically thin (Miller 2000), so
the differential rotation across it is small. Convection may enforce
rigid rotation; however, one-dimensional calculations of bursts show
that convection persists only for the first $\lesssim 1\ {\rm s}$ of the
burst when the energy production is dominated by temperature sensitive
and rapid helium burning reactions. In \S 4, we found that short
wavelength baroclinic instabilities may act to enforce rigid rotation
within the burning layers. Theoretical studies of detailed burst models
and further investigations of angular momentum transport mechanisms are
needed to determine whether a coherent pulse is possible once convection
has ceased.

To compare the predicted spin down with observed values, we assume
that some mechanism operates to enforce rigid rotation within the
burning layers, but that the burning layers remain decoupled from the
underlying ashes, as suggested by the results of \S 4. Figure
\ref{fig:summ2} shows the spin evolution of the atmosphere assuming
that the whole atmosphere rigidly rotates. For both convective and
fully-radiative models, we ignore any differential rotation, and
assume $I\Omega$ is constant, where $I$ is the moment of inertia of
the atmosphere (eq. [\ref{eq:I}]). As in Figure \ref{fig:summ}, the
evolution during the burst is shown by the arrows, first the
atmosphere spins down as flux increases, then spins back up as burning
proceeds, increasing the mean molecular weight, and as the atmosphere
cools. To compare with observations, we also indicate the observed
frequency shifts (Table \ref{tab:obs}). For those bursts in which the
oscillation frequency was seen only in the tail (Table \ref{tab:obs}),
we plot $\Delta\Omega/\Omega$ as a lower limit, since some cooling
will have occurred before the oscillation is first seen.

Figure \ref{fig:summ2} shows that the change in spin frequency for
pure He ignition models is less than that for mixed H/He ignition,
because of the smaller hydrostatic expansion. For the largest
frequency shifts observed, $\Delta\Omega/\Omega\approx 8\times
10^{-3}$, pure He ignition models require a base temperature very
close to the limiting value from radiation pressure
(eq. [\ref{eq:Tmax}]) and $F>F\edd$, so that the density at the base
is decreased, enhancing the expansion. It is much easier to get
$\Delta\Omega/\Omega$ values this large with mixed H/He ignition,
although as we describe below, all bursts with oscillations so far
have the characteristics of pure He bursts.

In \S \ref{sec:expansion}, we found that the time for heat to diffuse
from the burning layers to the photosphere is $\approx 1\ {\rm
s}$. This delay explains why the spin down of the atmosphere is not
observed at the beginning of a burst; the expansion occurs before the
signal is seen. We have also shown that the vertical propagation of
heat through the atmosphere can affect the oscillation amplitude. A
requirement for a large amplitude oscillation is that the time to
transport heat from the burning layers (or the top of the convection
zone) to the photosphere must be small compared to the shearing time
across the atmosphere $\approx \Delta\nu^{-1}$. If so, the shearing
atmosphere transmits the burning pattern from below, much like passing
a flashlight under a piece of paper. If not, the oscillation will be
smeared out, the atmosphere being heated uniformly by the quickly
revolving burning layers.

Our results show that there are important differences in spin
evolution depending on whether the helium ignites in a pure helium or
mixed H/He environment. Because of the greater mean molecular weight
of pure helium as opposed to a solar composition, the expansion of the
atmosphere in pure He bursts is roughly half that of mixed H/He
bursts, giving a smaller $\Delta\Omega/\Omega$ (Figure
\ref{fig:summ2}). Also, the ratio of thermal time to shearing time is
smaller for a pure helium atmosphere, and the energy production more
localized because of the temperature sensitivity of He burning
reactions, increasing the likelihood of observing coherent
oscillations for pure He bursts.

\subsection{Why Do Some Bursts Show Oscillations, But Not All?}\label{sec:muno}

A puzzle is that oscillations are not seen in bursts from all sources,
or in all bursts from a particular source. So far, there has been only
one study of the relation between burst oscillations and properties of
the bursts. Muno et al. (2000) studied nine X-ray bursts from KS
1731-26 which occurred during observations with ${\it RXTE}$ at
various times between 1996 July and 1999 February. They found that
bursts which show coherent oscillations are of short duration
($\approx 10\ {\rm s}$), show radius expansion, and have high peak
flux. These characteristics are typical of helium rich 
bursts. Bursts with longer durations typical of H burning showed no
oscillations. More studies are needed to see if this is a general
result for other sources. Certainly all reported burst oscillations
that we can find appear to be in bursts of short duration typical of
pure He ignition. As far as we are aware, oscillations have not been
detected in bursts of long duration ($>50 \ {\rm s}$) typical of H burning.

Why should only pure-He ignition bursts show oscillations? The
differences we find between pure He and mixed H/He bursts may provide an
explanation. As we described in \S 3.4 (see Figure \ref{fig:therm}), we
find that the thermal time is less and the shearing time is greater for
pure helium bursts, making it more likely that a large amplitude
oscillation from deep regions may propagate outwards.

\subsection{Long Term Stability of the Oscillation Frequency}\label{sec:recouple}

The frequency evolution reported for six bursts is well fit by the model
$\nu(t)=\nu_0-\Delta\nu e^{-t/\tau}$, where $\nu_0$ is the frequency in
the burst tail, $\Delta\nu$ the amount by which the oscillation
frequency changes and $\tau$ the decay time. The parameters of these
fits are given in Table 1. The observed decay times $\tau\approx 2$--$7\
{\rm s}$ are similar to the expected cooling time of the burning layers
(eq. [\ref{eq:ttherm}]). However, we found in \S 4 that there are
mechanisms that might couple the burning layers to the star on a similar
timescale. Can we distinguish between these two possibilities
observationally?

Figure \ref{fig:summ} shows that if the burning layers remain decoupled,
the frequency observed in the burst tail will not be that of the neutron
star spin. Because of the greater mean molecular weight of the ashes,
the thickness of the cooling atmosphere in the tail is different to the
initial thickness by $\approx 1\ {\rm m}$.  This change in thickness
depends on how complete the burning was, so that variations in the
energetics and burning in the burst would translate into a scatter of
$\approx 1$ part in $10^4$ in the final frequency.  If the oscillation
frequency could be shown to be more stable than this from one burst to
the next, it would imply that the ashes and burning layers must recouple
during the burst decay. This is complicated due to Doppler shifts from
the orbital motion of the neutron star, which change the observed
frequency at a level $\approx 2\times 10^{-3}(v_{\rm orb}/600\ {\rm km\
s^{-1}}$) across the orbit. Strohmayer et al. (1998b) showed that two
bursts from 4U 1636-54 showed a frequency difference consistent with
orbital Doppler shifts, while two bursts from 4U 1728-34 separated by
more than one year showed the same asymptotic frequency to $\approx 1$
part in $10^4$ (see Table 1). If the orbital parameters are known and
the orbital Doppler shift accounted for, more measurements such as these
would indicate whether recoupling occurs in the burst tail.

\section{Conclusions}

We have shown that the hydrostatic expansion, spin down, and later spin
up during contraction of the neutron star atmosphere naturally explains
the magnitude and sign of the observed changes in the nearly coherent
oscillations observed in Type I X-ray bursts. Our results support
the simple picture proposed by Strohmayer et al. (1997a) to explain the
frequency evolution seen in Type I bursts, and thus the identification
of the burst oscillation frequency with the neutron star spin.

The amplitude of the oscillation is set not only by the lateral extent
of the asymmetry on the neutron star surface, but also by the vertical
propagation of heat through the shearing atmosphere. In addition, we
find that the spin behavior differs depending on whether the burst
results from pure He ignition or He ignition in a H-rich
environment. The spin down is smaller by roughly a factor of two for
pure He ignition as opposed to mixed H/He ignition, because the pure He
models expand outwards less due to the larger mean molecular weight. In
addition, the time for heat propagation through the atmosphere is
smaller for pure He models.

This might explain why oscillations have not been detected during bursts
of long duration ($\gtrsim 30\ {\rm s}$) typical of H burning (for
example, the bursts from GS 1826-24 shown in Figures 2 and 5 of Kong et
al. 2000). In particular, the recent study of Muno et al. (2000) of
bursts from KS 1731-26 found oscillations only during those bursts with
short durations $\approx 10\ {\rm s}$, typical of pure He
ignition. However, the largest values of $\Delta\Omega/\Omega$ that are
observed (for 4U~1702-43 and 4U~1728-34) require base temperatures very
close to the limiting value from radiation pressure for pure He bursts,
and this may prove to be a problem. More observational studies of the
phenomenology of burst oscillations and their relation to burst
properties and source properties, in particular accretion rate, are
needed. With further theoretical work, these studies could give us
important clues as to how the nature of nuclear burning during bursts
depends on accretion rate, and help to resolve the disagreement with
one-dimensional models first found with EXOSAT (Bildsten 2000 and
references therein).

We have not found a robust hydrodynamical mechanism that recouples the
burning layers to the underlying star in the $\approx 10\ {\rm s}$ burst
duration.  However, the coherence of the observed burst oscillations
suggests that little differential rotation occurs within the burning
layer itself. As we note in \S \ref{sec:coupling}, this might be
accomplished through short vertical wavelength baroclinic instabilities,
though more theoretical work is needed.

\begin{figure}
\plottwo{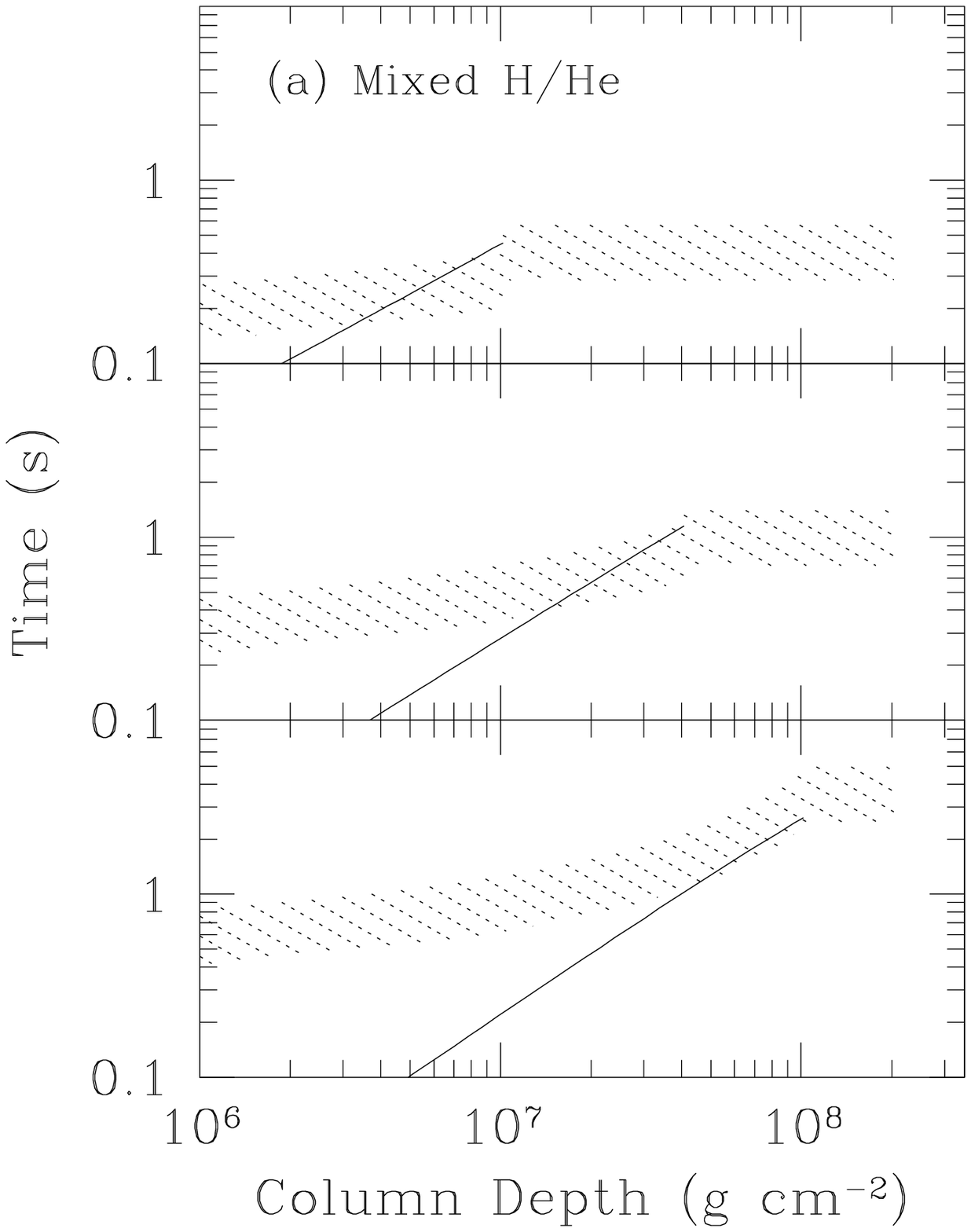}{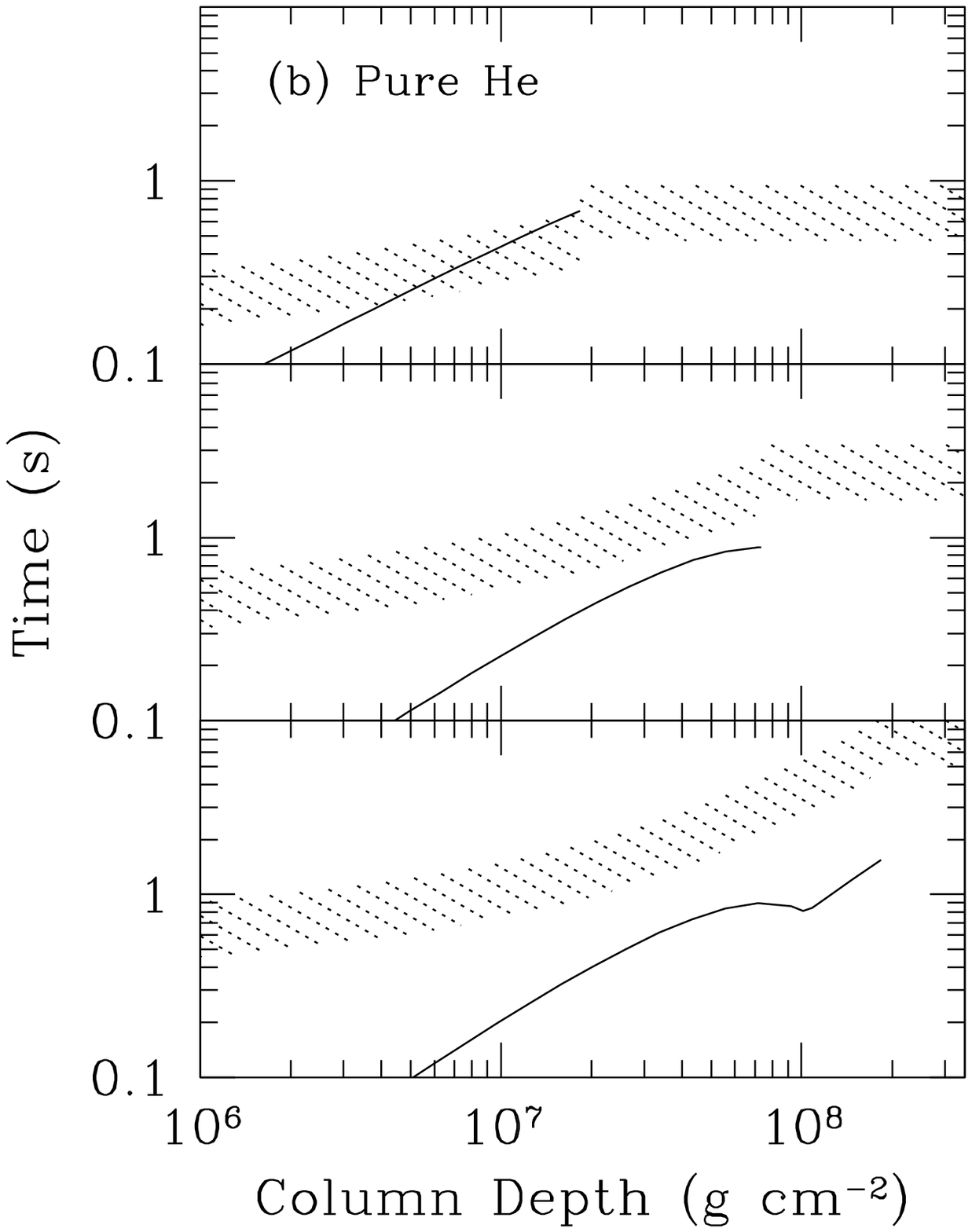}
\caption{Thermal time $t_{\rm therm}$ (eq.[\ref{eq:ttherm}]) and wrap
around time $\Delta\nu^{-1}$ for the convective models of Table
\ref{tab:conv}. In each case, the solid line shows $t_{\rm therm}$,
and the hatched region $\Delta\nu^{-1}$ for $\nu_s$=300--600 Hz. (a)
Mixed H/He ignition convective models with (top to bottom)
$F/F\edd=1.15, 0.74$ and $0.31$. (b) Pure He ignition convective
models with (top to bottom) $F/F\edd=1.21, 0.59$ and $0.35$.
\label{fig:therm}}
\end{figure}

\begin{figure}
\plottwo{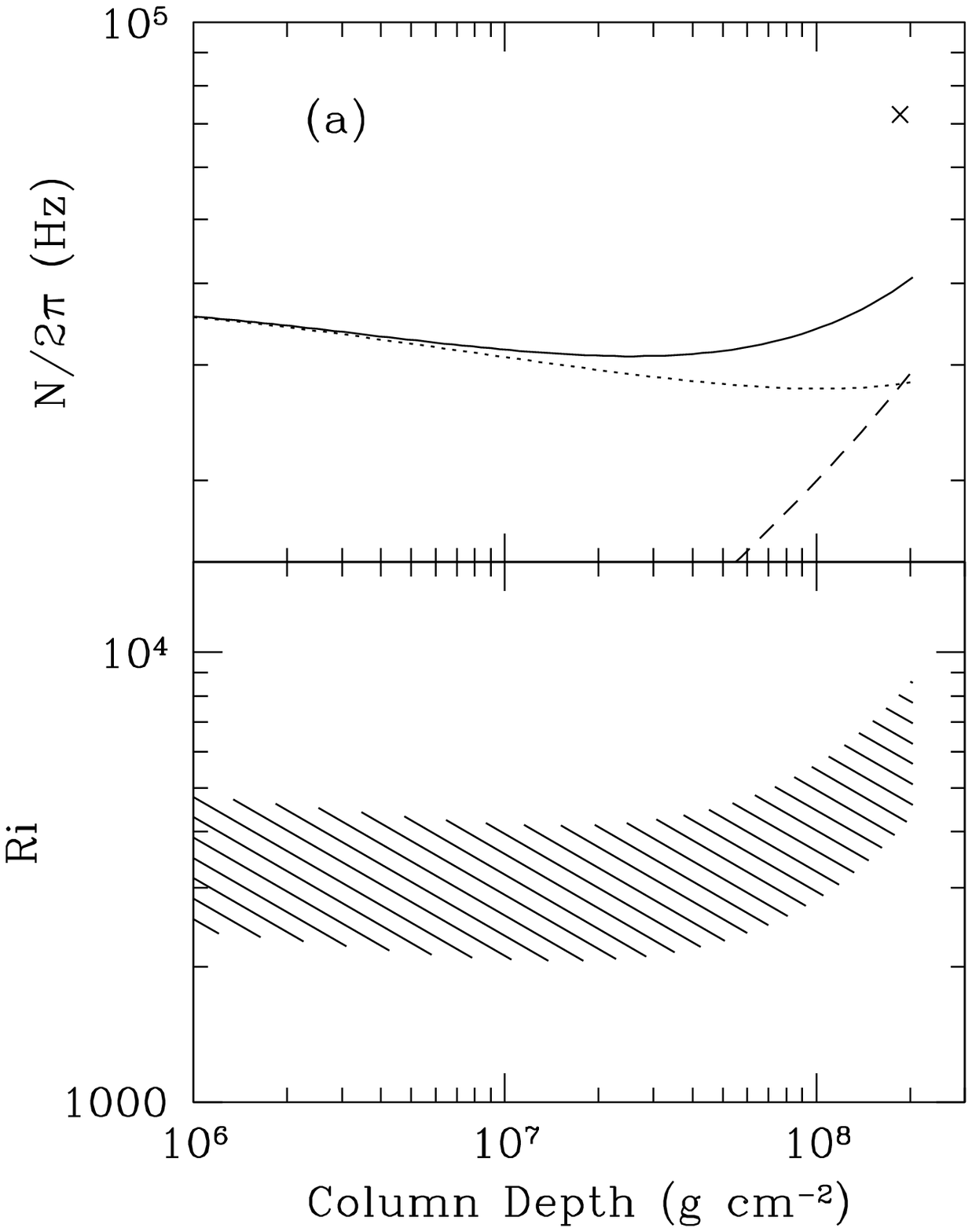}{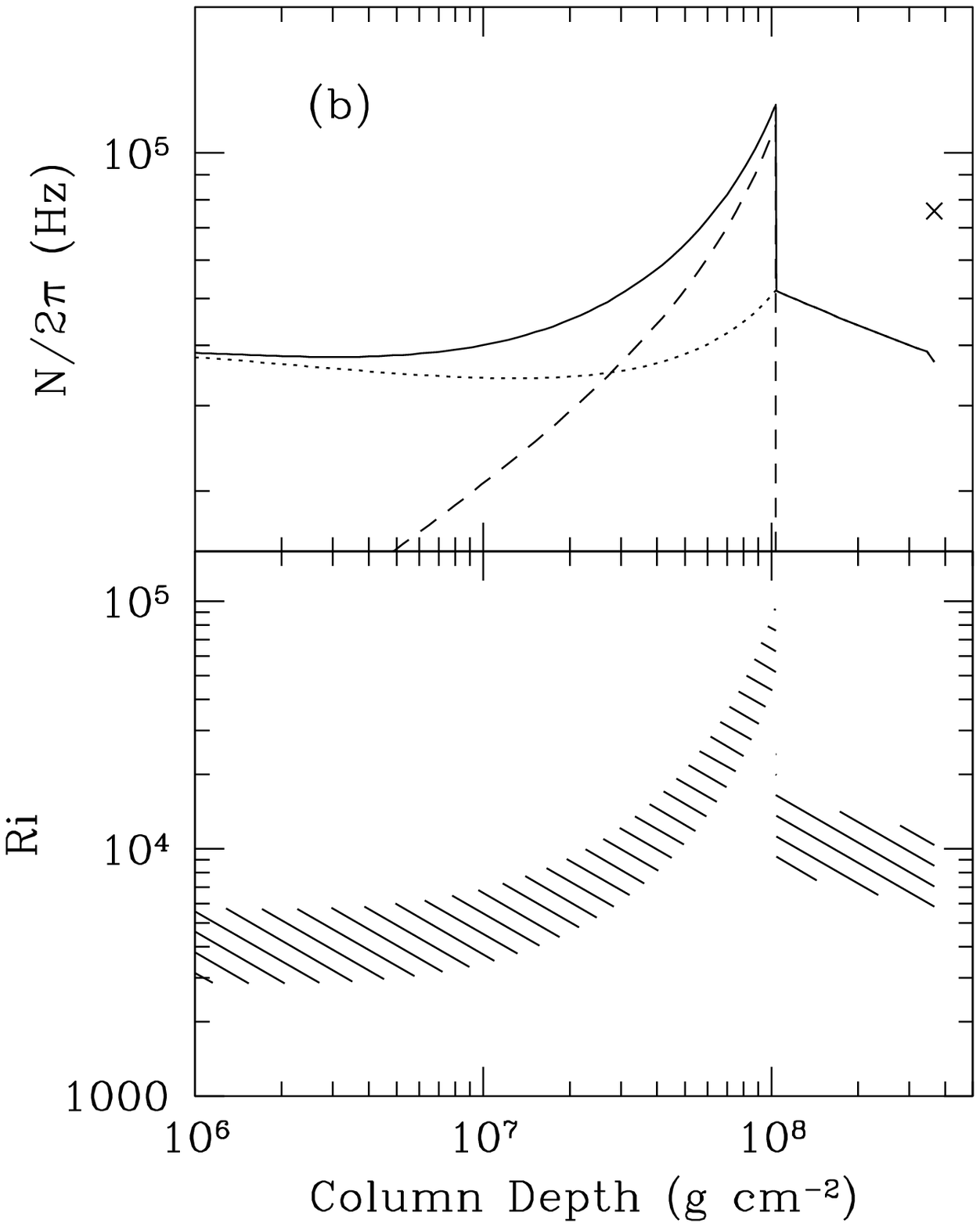}
\caption{\BV frequency $N/2\pi$ and Richardson number ${\rm Ri}$ for
(a) the fully-radiative, constant $F=0.74\ F\edd$ mixed H/He ignition
model and (b) the fully-radiative, constant $F=0.59\ F\edd$ pure He
ignition model. We show the total buoyancy by a solid line, the
thermal buoyancy as a dotted line and the composition piece as a
dashed line. The cross shows an estimate of $N/2\pi$ at the base,
$N^2_{\rm base}=(g/H_b)(\Delta\ln\mu)$, where we take the ashes as
having $\mu=A/(1+Z)=2.1$. For the pure He ignition model (b), there is
a peak in the buoyancy at the place where the hydrogen runs out
($y\approx y_d$), at this depth the composition piece of the buoyancy
dominates the thermal piece.
\label{fig:ri1}}
\end{figure}

\begin{figure}
\plottwo{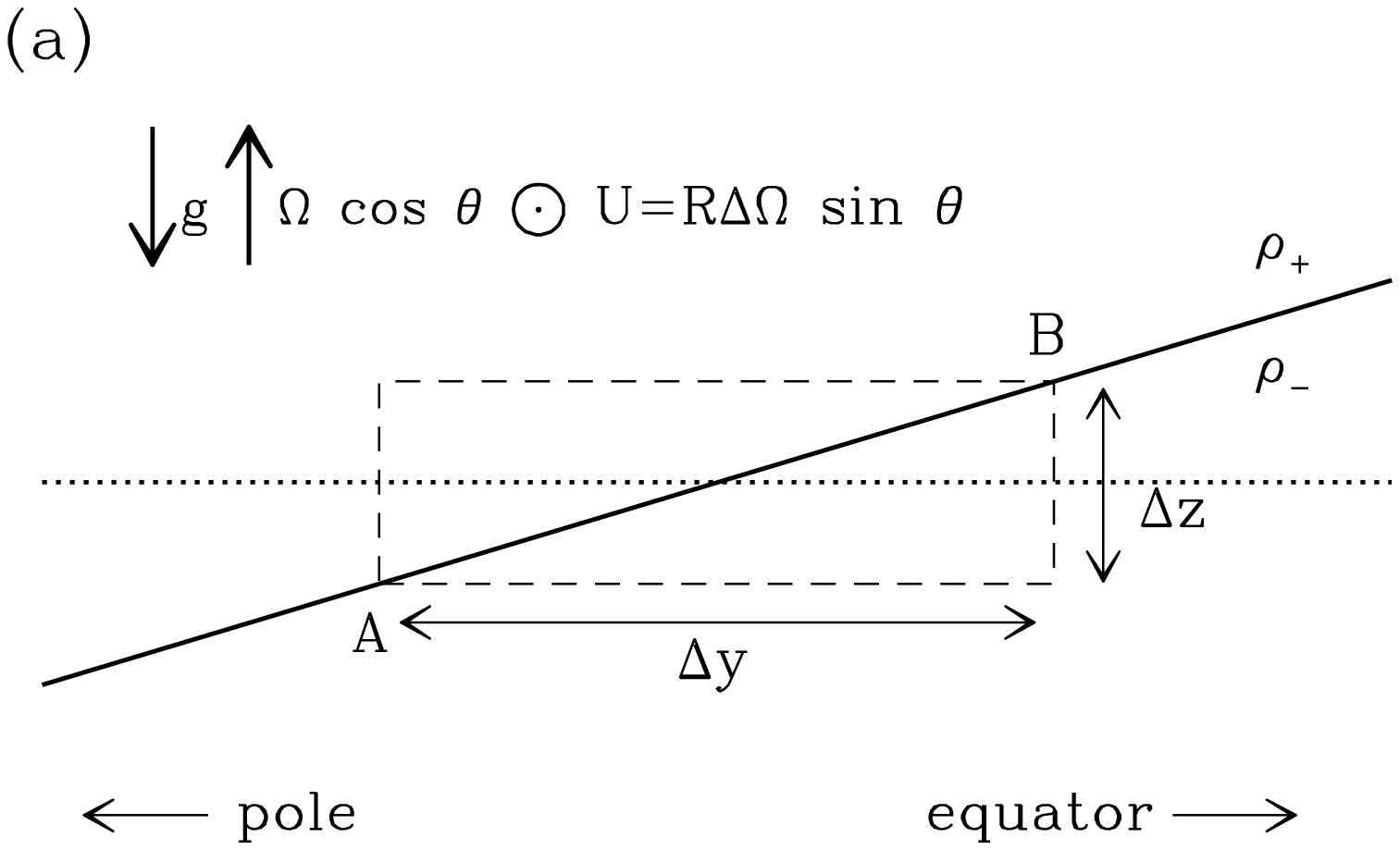}{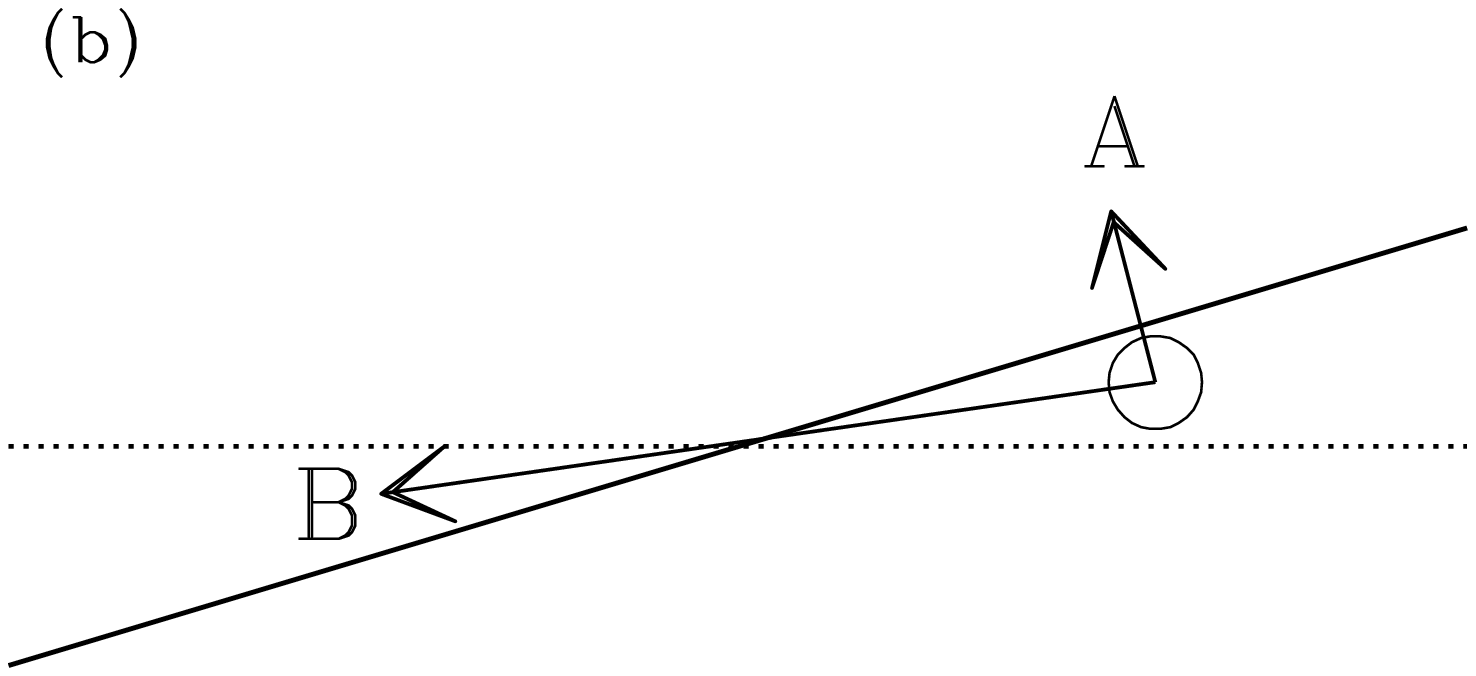}
\caption{A section of the two-layer model. The burning layers and
ashes are represented by layers of different constant density. The
radial coordinate $z$ increases upwards, the transverse co-ordinate
$y$ increases to the right. The vertical component of the rotation
vector is $\Omega\cos\theta$, and the upper fluid (density
$\rho_+<\rho_-$) is moving out of the page with velocity
$R\Delta\Omega\sin\theta$. (a) The boundary between the two fluids
(solid line) slopes because in the upper fluid there is horizontal
pressure gradient which balances the Coriolis force (see text).  (b)
The nature of the baroclinic instability. If the fluid element is
moved to point A, it is heavier than its surroundings, and experiences
a restoring force. If it is moved to point B, however, it is heavier
than its new surroundings but has dropped in the gravitational field,
releasing energy. All motions within the so-called ``wedge of
instability'' are convectively unstable in this way.
\label{fig:slope}} 
\end{figure}

\begin{figure}
\epsscale{0.9}\plotone{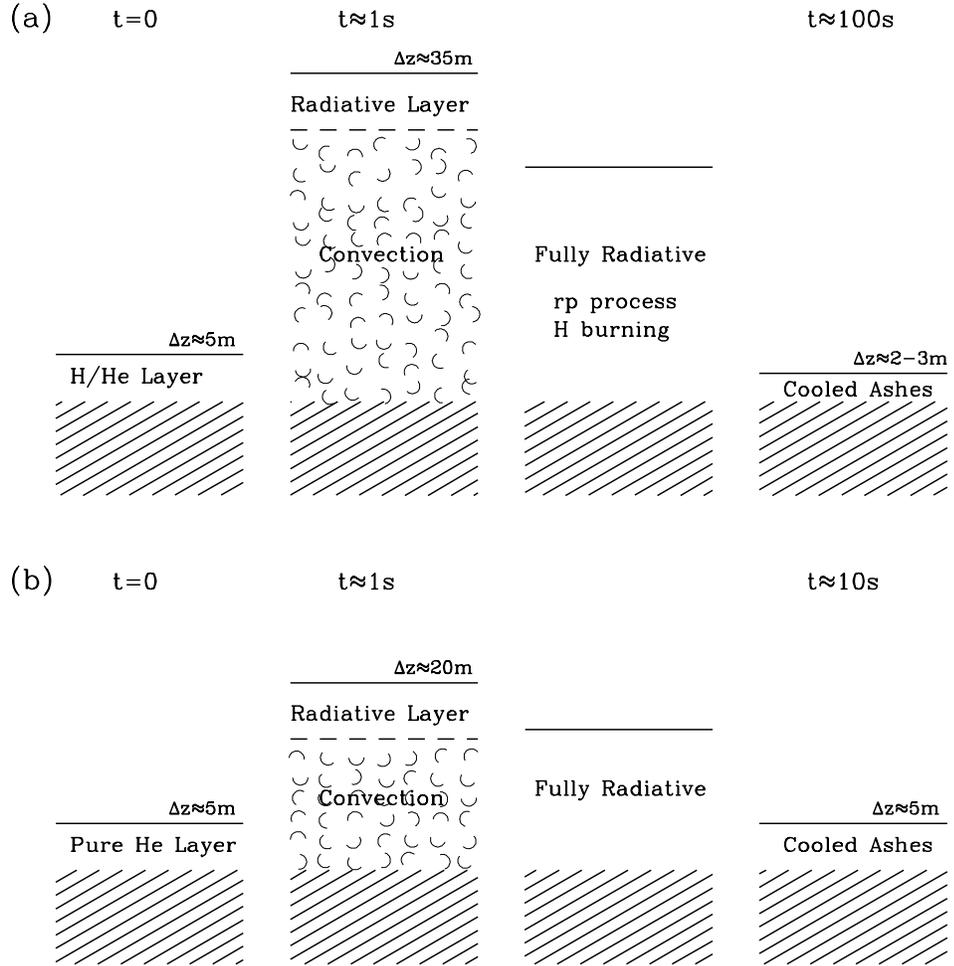}\epsscale{1.0}
\caption{Schematic of the evolution of the atmosphere during a burst
for (a) mixed H/He ignition and (b) pure He ignition. The rapid energy
release from helium burning reactions makes the initial stages of many
bursts convective. After a time $\lesssim 1\ {\rm s}$ (see text for
references to simulations of bursts), the convection zone shrinks and
disappears, leaving a radiative atmosphere. For pure He ignition, the
burst duration is typically $\approx 10\ {\rm s}$; for mixed H/He
ignition, rp process H burning in the tail can prolong the burst for a
further $\gtrsim 30\ {\rm s}$.\label{fig:scheme}}
\end{figure}

\begin{figure}
\epsscale{0.6}\plotone{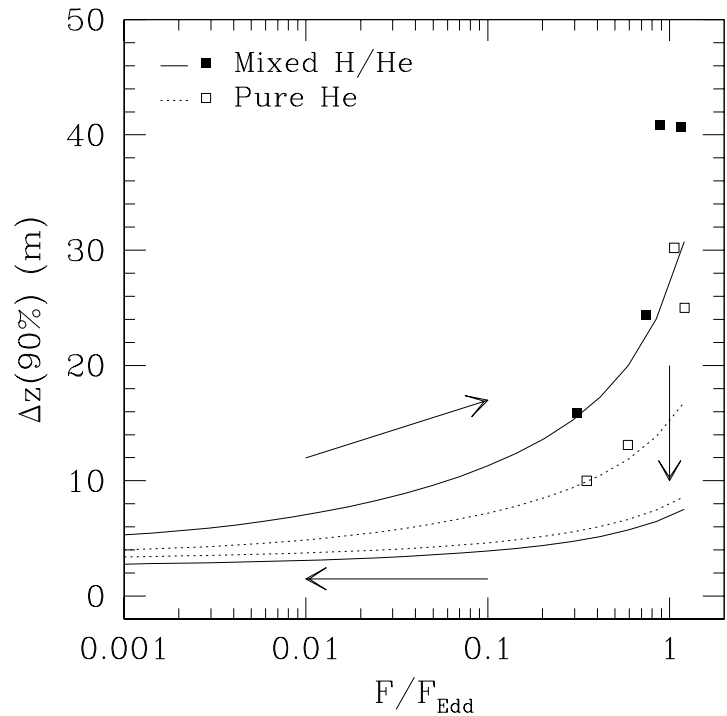}\epsscale{1.0}
\caption{The thickness of the atmosphere which contains 90\% of the
mass, $\Delta z(90$\%$)$, as a function of the flux for constant flux
radiative atmospheres (solid and dotted lines) and for the convective
models of Table \ref{tab:conv} (solid and open squares). For the
radiative models, the upper curve is for composition profile the same
as at ignition, while the lower curve is for a composition of
$^{73}$Kr ($^{56}$Ni) for the mixed H/He (pure He) case, chosen to
represent the products of burning. Initially, the evolution of the
thickness of the layer is along or above the upper curve as it ignites
and heats up, depending on the extent of the convection zone. As
nuclear burning proceeds, the mean molecular weight increases, and the
thickness decreases, eventually moving back along the lower curve as
the atmosphere cools. Just before ignition (upper curve, $F\ll
F\edd$), the thickness of the accumulated layer is $\approx 5\ {\rm
m}$. During the burst, the atmosphere expands hydrostatically by
$\Delta z\approx 10$--$40\ {\rm m}$ ($\approx 5$--$20\ {\rm m}$). In
the cooling tail, the thickness may differ by $\approx 1\ {\rm m}$
from that before the burst.\label{fig:summ}}
\end{figure}

\begin{figure}
\epsscale{1.0}\plotone{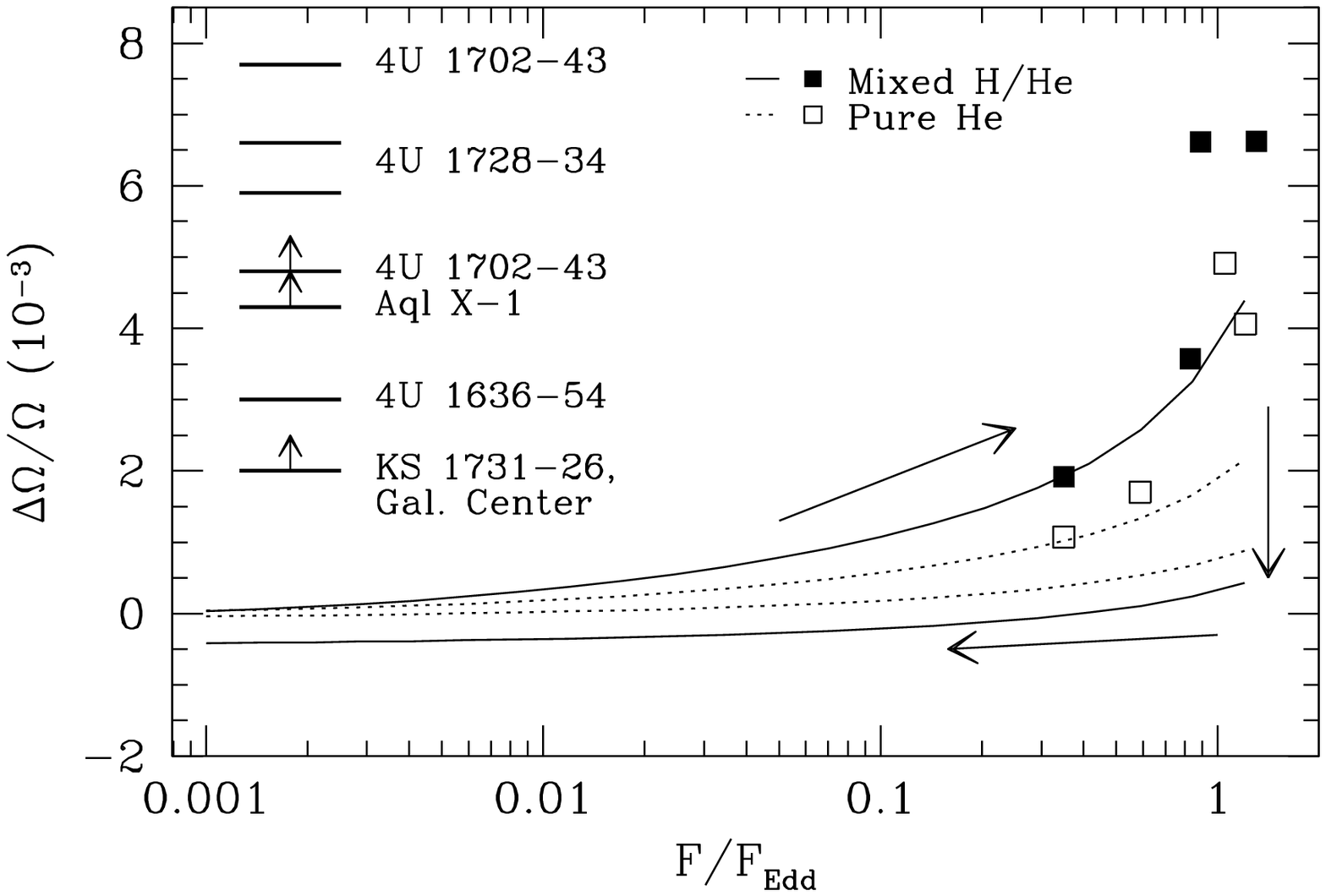}\epsscale{1.0}
\caption{The spin evolution of the atmosphere for convective (squares)
and radiative (lines) models, assuming rigid rotation is maintained
throughout the atmosphere. We plot the fractional spin frequency
change of the burning shell, $\Delta\Omega/\Omega$, such that
$\Delta\Omega/\Omega>0$ indicates spin down. As in Figure
\ref{fig:summ}, the evolution during the burst is as shown by the
arrows. We also indicate the observed frequency shifts for those
objects listed in Table \ref{tab:obs} by horizontal bars. For those
bursts in which the oscillation frequency was seen only in the tail,
we plot the $\Delta\Omega/\Omega$ value as a lower
limit.\label{fig:summ2}}
\end{figure}

We have not discussed the complex question of the lateral spreading of
the burning front during the burst rise, or the cause of asymmetry in
the cooling tail of bursts. In the simplest picture of a spreading
burning front (Strohmayer et al. 1997b; Strohmayer et al. 1998c) it is a
mystery why oscillations are seen in the burst tail, when the whole
surface has ignited. One possible explanation is that an asymmetry
arises because the atmosphere takes a finite time to cool once ignited,
and a finite time to spread around the star. Thus the matter on the
opposite side of the star from the ignition point will be a little
hotter since it ignited later. If the flux at each point is $\propto
e^{-t/\tau}$ once ignited, and the burning front propagates around the
whole star (starting at the equator) in a time $t_{\rm burn}$, it is
straightforward to show that the peak-to-peak luminosity amplitude in
the tail for an observer looking down on the ignition point is $\Delta
L/L=2(1-\exp(-t_{\rm burn}/\tau))/(1+\exp(-t_{\rm burn}/2\tau))^2$, or,
for $t_{\rm burn}\ll\tau$, $\Delta L/L\approx t_{\rm burn}/2\tau$. Thus
for a spreading time $t_{\rm burn}\approx 0.5\ {\rm s}$ and $\tau\approx
10\ {\rm s}$, the peak-to-peak amplitude in the tail is a few percent, a
little less than the 5--20\% amplitudes observed (Smith, Morgan \& Bradt
1997; Zhang et al. 1998; Strohmayer et al. 1998a). Another possibility
is that the oscillations come and go because of changes in the vertical
transport of heat rather than changes in the size of the asymmetry on
the stellar surface. For example, Figure \ref{fig:therm} shows that as
the atmosphere heats up during the burst rise, the ratio of thermal time
to wrap-around time $t_{\rm therm}\Delta\nu$ increases. As this ratio
increases, it becomes more likely that the oscillation will be washed
out. It may be that at the burst peak, the oscillations disappear
because of this effect.

We have not addressed what mechanism might cause frequency doubling
during bursts, as seems to be the case for at least one source (4U
1636-56; Miller 1999). There are two processes that we have discussed
in this paper that may play a role. The first is the recoupling
mechanism. If the burning layers recouple to the neutron star during
the burst, they may do so in an $m=2$ pattern, leading to frequency
doubling. Our linear instability analysis of the two-layer beta-plane
model (\S \ref{sec:baroclinic}) suggests that the baroclinic
instability would have $m\approx 20$ when it goes unstable. However,
we have not calculated the correct spherical eigenfunctions. The
second possible process is the azimuthal wrapping of the the magnetic
field. If an initially dipolar field is wound up, an $m=2$ pattern
might emerge in the flow.

A test for the rotational modulation hypothesis might be the
distribution of observed $\Delta\Omega/\Omega$ values. For sources
undergoing X-ray bursts of the same type (for example, similar accretion
rates and same type of ignition) on similar mass neutron stars, we would
expect the distribution of $\Delta\Omega/\Omega$ values to be the same
for each source, and independent of the spin frequency or number of
hotspots on the surface. In fact, the distribution of
$\Delta\Omega/\Omega$ so far (Table 1) looks bimodal, though we should
be wary of small number statistics. Two sources, 4U 1728 and 4U 1702,
which have $\nu_0\approx 350\ {\rm Hz}$ show frequency changes
$\Delta\nu/\nu\approx 5$--$8\times 10^{-3}$, whereas the other four
objects which have $\nu_0\approx 550\ {\rm Hz}$ show frequency changes
about half as large, $\Delta\nu/\nu\approx 1$--$4\times 10^{-3}$. If
further observations show this bimodal distribution to be true, this
might indicate that the expansion of the atmosphere is less for the
$\nu_0\approx 550\ {\rm Hz}$ objects, for example, because of pure He
rather then mixed H/He ignition. Alternatively, we note that the
$\Delta\nu$ values are very similar for all six objects, and this might
point to another explanation for the observed frequency evolution. Care
must be taken that the observed $\Delta\Omega/\Omega$ represents the
change in burst oscillation frequency from the burst rise to the burst
tail. For example, if the oscillation is seen only during the burst
tail, the observed $\Delta\Omega/\Omega$ does not represent the full
spin down of the atmosphere, as in this case some cooling will have
occurred before the oscillation is seen.

We have not discussed other explanations for the frequency drifts. One
possibility is that the burst oscillation is a non-radial oscillation
(NRO) on the neutron star surface (Livio \& Bath 1982; McDermott \& Taam
1987; Bildsten \& Cutler 1995; Strohmayer \& Lee 1996). Non-radial
oscillations of the frequency observed would be plausible on a rapidly
rotating star, as the Coriolis force takes the typically much
lower-frequency modes and makes them mugh higher (Bildsten, Ushomirsky
and Cutler 1996). The observed frequency would be close to the spin
frequency of the star or some multiple $m\Omega$, and might change by a
$1$--$2\ {\rm Hz}$ during the burst as the atmosphere cooled. However,
it is not clear why only a single mode would be selected, and why that
frequency would be stable on timescales of a year or more, for which
conditions in the surface layers of the star would most likely change
significantly.

The physics of expansion-induced shear during shell flashes may prove
important in other contexts. For example, mixing due to shear
instabilities during a classical nova explosion could provide the
enrichment of CNO elements necessary to achieve a fast nova (see Truran
1982 for a review of the enrichment problem). Previous authors have
investigated shear mixing during accumulation of the fuel (see Livio \&
Truran 1990 for a review), but not during the burst itself when shear is
induced by expansion of the atmosphere as in a Type I X-ray burst. A
similar situation arises in AGB stars during He shell flashes, where
mixing of carbon into the proton rich envelope is necessary to allow
production of s-process elements (Iben 1991; Sackmann \& Boothroyd 1991;
Langer et al. 1999), and perhaps could be achieved by shear
instabilities during shell flashes.

\acknowledgements We thank Derek Fox, Daniel Holz, Yuri Levin, Mike
Muno, Tod Strohmayer, Chris Thompson, Greg Ushomirsky, Marten van
Kerkwijk and Ellen Zweibel for valuable discussions and comments. We
thank the Astronomical Institute, ``Anton Pannekoek'' of the
University of Amsterdam for hospitality and for supporting L. B.'s
visit there as the CHEAF Visiting Professor. This research was
supported by NASA via grants NAG 5-8658 and NAGW-4517 and by the
National Science Foundation under Grants No. PHY94-07194 and
AST97-31632. L. B. is a Cottrell Scholar of the Research Corporation.


\begin{references}

\noindent
Ayasli, S., \& Joss, P. C. 1982, \apj, 256, 637 

\noindent
Benton, E. R., \& Clark, A. 1974, Ann. Rev. Fluid Mech., 6, 257

\noindent
Bhattacharya, D. 1995 in X-Ray Binaries, ed. W. H. G. Lewin, J. van
Paradijs, \& E. P. J. van den Heuvel (Cambridge: Cambridge University
Press), 233

\noindent
Bildsten, L. 1995, \apj, 438, 852

\noindent
Bildsten, L. 1998, in The Many Faces of Neutron Stars,
ed. R. Buccheri, J. van Paradijs, \& M. A. Alpar (Dordrecht: Kluwer), 419

\noindent
Bildsten, L. 2000, in Cosmic Explosions, ed. S. S. Holt and
W. W. Zhang (New York:AIP)

\noindent
Bildsten, L., \& Brown, E. F. 1997, \apj, 477, 897

\noindent
Bildsten, L., \& Cumming, A. 1998, \apj, 506, 842

\noindent
Bildsten, L., \& Cutler, C. 1995, \apj, 449, 800

\noindent
Bildsten, L., Ushomirsky, G., \& Cutler, C. 1996, \apj, 460, 827

\noindent
Brekhovskikh, L. M., \& Goncharov, V. 1994, Mechanics of Continua and Wave Dynamics (New York: Springer)

\noindent
Brown, E. F.,  2000, \apj, 531, 988 

\noindent
Brown, E. F., \& Bildsten, L. 1998 \apj, 496, 915

\noindent
Chakrabarty, D., \& Morgan, E. H. 1998, \nat, 394, 346 

\noindent
Chandrasekhar, S. 1961, Hydrodynamic and Hydromagnetic Stability (New
York:Dover)

\noindent
Clayton, D. D. 1983, Principles of Stellar Evolution and
Nucleosynthesis (Chicago: The University of Chicago Press)

\noindent
Cox, J. P., \& Giuli, R. T. 1968, Principles of Stellar Structure (New York:
Gordon \& Breach)

\noindent
Dutton, J. A. 1995, Dynamics of Atmospheric Motion (New York: Dover)

\noindent
Endal, A. S., \& Sofia, S. 1978, \apj, 220, 279

\noindent
Fox, D. W., Muno, M. P., Morgan, E. H., \& Bildsten L. 2000, in preparation

\noindent
Fryxell, B. A., \& Woosley, S. E. 1982, \apj, 261, 332

\noindent
Fujimoto, M. Y. 1988, \aap, 198, 163

\noindent
Fujimoto, M. Y. 1993, \apj, 419, 768

\noindent
Fujimoto, M. Y., Hanawa, T., \& Miyaji, S. 1981, \apj, 247, 267 (FHM)

\noindent
Fujimoto, M. Y., Sztajno, M., Lewin, W. H. G., \& van Paradijs,
J. 1987, \apj, 319, 902

\noindent
Fushiki, I., \& Lamb, D. Q. 1987a, \apj, 317, 368

\noindent
Fushiki, I., \& Lamb, D. Q. 1987b, \apj, 323, L55

\noindent
Greenspan, H. P., \& Howard, L. N. 1963, J. Fluid Mech., 17, 385

\noindent
Hanawa, T., \& Fujimoto, M. Y. 1982, \pasj, 34, 495

\noindent
Hanawa, T., \& Fujimoto, M. Y. 1984, \pasj, 36, 199

\noindent
Hanawa, T., \& Sugimoto, D. 1982, \pasj, 34, 1

\noindent
Hollingsworth, A. 1975, Quart. J. Roy. Meteorol. Soc., 101, 495

\noindent
Hollingsworth, A., Simmons, A. J., \& Hoskins, B. J. 1976, Quart. J. Roy. Meteo
rol. Soc., 102, 901

\noindent
Holton, J. R. 1965, J. Atmos. Sci., 22, 402

\noindent
Hoyle, R., \& Fowler, W. A. 1965, in Quasi-Stellar Sources and
Gravitational Collapse, ed. I. Robinson, A. Schild, \& E. L. Shucking
(Chicago: University of Chicago Press)

\noindent
Iben, I. Jr., 1991, in Evolution of Stars: The
Photospheric Abundance Connection, ed. G. Michaud \& A. Tutukov, 257

\noindent
Joss, P. C. 1977, Nature, 270, 310 

\noindent
Joss, P. C. 1978, \apj, 225, L123 

\noindent
Joss, P. C., \& Li, F. K. 1980, \apj, 238, 287

\noindent
Knobloch, E., \& Spruit, H. C. 1982, \aap, 113, 261

\noindent
Koike, O., Hashimoto, M., Arai, K., \& Wanajo, S. 1999, \aap, 342, 464

\noindent
Kong, A. K. H., Homer, L., Kuulkers, E., Charles, P. A., \& Smale,
A. P. 2000, \mnras, 311, 405

\noindent
Langer, N., Heger, A., Wellstein, S., \& Herwig, F. 1999, \aap, 346,
L37

\noindent
Lewin, W. H. G., van Paradijs, J., \& Taam, R. E. 1995, in X-Ray
Binaries, ed. W. H. G. Lewin, J. van Paradijs, \& E. P. J. van den
Heuvel (Cambridge: Cambridge University Press), 175

\noindent
Livio, M., \& Bath, G. T. 1982, \aap, 116, 286

\noindent
Livio, M., \& Truran, J. W. 1987, \apj, 318, 316

\noindent
Livio, M., \& Truran, J. W. 1990, Ann. N. Y. Acad. Sci., 617, 126

\noindent
McDermott, P. N., \& Taam, R. E. 1987, \apj, 318, 278

\noindent
Miller, M. C. 1999, \apj, 515, L77

\noindent
Miller, M. C. 2000, \apj, 531, 458 

\noindent
Moura, A. D., \& Stone, P. H. 1976, J. Atmos. Sci., 33, 602

\noindent
Muno, M. P., Fox, D. W., Morgan, E. H., \& Bildsten L. 2000, Ap J, to appear

\noindent
Nozakura, T., Ikeuchi, S., \& Fujimoto, M. Y. 1984, \apj, 286, 221 

\noindent
Paczynski, B. \& Anderson, N. 1986, \apj, 302, 1 

\noindent
Pedlosky, J. 1987, Geophysical Fluid Dynamics (New York: Springer)

\noindent
Psaltis, D., \& Chakrabarty, D. 1999, \apj, 521, 332

\noindent
Psaltis, D., \& Lamb, F. K. 1998, in Neutron Stars and Pulsars,
ed. N. Shibazaki, N. Kawai, S. Shibata, \& T. Kifune (Tokyo:
Univ. Acad. Press), 179

\noindent
Sackmann, I.-J., \& Boothroyd, A. I. 1991, in Evolution of Stars: The
Photospheric Abundance Connection, ed. G. Michaud \& A. Tutukov, 275

\noindent
Sakurai, T., Clark, A. Jr., \& Clark, P. A. 1971, J. Fluid Mech., 49, 753

\noindent
Schatz, H., et al. 1998, Phys. Rep., 294, 167

\noindent
Schatz, H., Bildsten, L., Cumming, A., \& Wiescher, M. 1999, \apj, 524, 1014 

\noindent
Schoelkopf, R. J., \& Kelley, R. L. 1991, \apj, 375, 696

\noindent
Shapiro, S. L., \& Teukolsky, S. A. 1983, Black Holes, White Dwarfs
and Neutron Stars: The Physics of Compact Objects (New York:Wiley)

\noindent
Shara, M. M. 1982, \apj, 261, 649

\noindent
Simmons, A. J., \& Hoskins, B. J. 1976, J. Atmos. Sci., 33, 1454

\noindent
Simmons, A. J., \& Hoskins, B. J. 1977, J. Atmos. Sci., 34, 581

\noindent
Smith, D. A., Morgan, E. H., \& Bradt, H. 1997, \apj, 479, L137

\noindent
Spitzer, L. Jr. 1962, Physics of Fully Ionized Gases (2nd Ed.; New York: Wiley)

\noindent
Spruit, H. C. 1999, \aap, 349, 189

\noindent
Spruit, H. C., \& Knobloch, E. 1984, \aap, 132, 89

\noindent
Strohmayer, T. E. 1999a, \apj, 523, L51 

\noindent
Strohmayer, T. E. 1999b, preprint (astro-ph/9911338)

\noindent
Strohmayer, T. E., Jahoda, K., Giles, B. A., \& Lee, U. 1997a, \apj, 486, 355

\noindent
Strohmayer, T. E., \& Lee, U. 1996, \apj, 467, 773 

\noindent
Strohmayer, T. E., \& Markwardt, C. B. 1999, \apj, 516, L81

\noindent
Strohmayer, T. E., Swank, J. H., \& Zhang, W. 1998a, Nucl. Phys. B
(Proc. Suppl.), 69(1-3), 129

\noindent
Strohmayer, T. E., Zhang, W., \& Swank, J. H. 1997b, \apj, 487, L77

\noindent
Strohmayer, T. E., Zhang, W., Swank, J. H., \& Lapidus, I. 1998b, \apj, 503, L147

\noindent
Strohmayer, T. E., Zhang, W., Swank, J. H., Smale, A., Titarchuk, L., Day, C., \& Lee, U. 1996, \apj, 469, L9

\noindent
Strohmayer, T. E., Zhang, W., Swank, J. H., White, N. E., \& Lapidus, I. 1998c, \apj, 498, L135

\noindent
Taam, R. E. 1980, \apj, 241, 358

\noindent
Taam, R. E., \& Picklum, R. E. 1978, \apj, 224, 210 

\noindent
Tassoul, J.-L., \& Tassoul, M. 1982, \apjs, 49, 317

\noindent
Truran, J. W. 1982, in Essays in Nuclear Astrophysics,
ed. C. A. Barnes, D. D. Clayton, \& D. Schramm (Cambridge: Cambridge
University Press), 467

\noindent
van der Klis, M. 2000, \araa, to appear September 2000 (astro-ph/0001167)

\noindent
Walin, G. 1969, J. Fluid Mech., 36, 289

\noindent
Wallace, R. K., \& Woosley, S. E. 1981, \apjs, 45, 389

\noindent
Wallace, R. K., Woosley, S. E., \& Weaver, T. A. 1982, \apj, 258, 696 (W82)

\noindent
Warn, H. 1976, J. Atmos. Sci., 33, 1478

\noindent
Wijnands, R., \& van Der Klis, M. 1998, \nat, 394, 344 

\noindent
Zhang, W., Jahoda, K., Kelley, R. L., Strohmayer, T. E., Swank, J. H., \& Zhang, S. N. 1998, \apj, 495, L9

\end{references}
\end{document}